\documentclass[32pt, usenatbib]{mn2e}
 \usepackage{graphicx,times,subfigure}
\usepackage{bm}
\usepackage{epsf}
\usepackage{amsmath}
\usepackage{amssymb}
\usepackage{cases}
\usepackage{float}
\usepackage{colortbl}
\usepackage{color}
\usepackage{multirow}
\usepackage{url}
\usepackage{hyperref}
\usepackage[utf8x]{inputenc} 

\def\la{\langle}
\def\ra{\rangle}
\def\n{\noindent}
\def\be{\begin{equation}}
\def\ee{\end{equation}}
\def\ben{\begin{eqnarray}}
\def\een{\end{eqnarray}}
\def\nn{\nonumber}

\def\myC{{\cal C}}
\def\myB{{\cal B}}
\def\myf{\kappa}
\def\bk{{\bf k}}

\def\myC{{\cal C}}
\def\myB{{\cal B}}

\def\bk{{\bf k}}

\def\2p{{(2\pi)^2}}

\def\be{\begin{equation}}
\def\ee{\end{equation}}
\def\beq{\begin{equation}}
\def\eeq{\end{equation}}
\def\ben{\begin{eqnarray}}
\def\een{\end{eqnarray}}
\def\bes{\begin{subequations}}
\def\ees{\end{subequations}}

\def\nn{{\nonumber}}

\newcommand{\beqa}{\begin{eqnarray}}
\newcommand{\eeqa}{\end{eqnarray}}

\def\ikap0{{\cal J}_{\theta_0}(r)}

\def\one1{\langle \kappa_{(i)}\kappa_{(j)} \rangle}
\def\one{{[\bar \xi^{(ij)}]}}

\def\ba{\begin{eqnarray}}
\def\ea{\end{eqnarray}}

\def\bk{{\bf k}}

\def\n{\noindent}

\def\myC{{\cal C}}

\def\bk{{\bf k}}

\def\2p{{(2\pi)^2}}

\def\be{\begin{equation}}
\def\ee{\end{equation}}

\def\beq{\begin{equation}}
\def\eeq{\end{equation}}

\def\ben{\begin{eqnarray}}
\def\een{\end{eqnarray}}

\def\nn{{\nonumber}}

\def\n{\noindent}

\def\myC{{{\cal C}}}

\def\myB{{\cal B}}

\def\bk{{\bf k}}

\def\2p{{(2\pi)^2}}

\def\btheta{{\boldsymbol\theta}}

\def\essA{{S^{(0)}_{\ell}}}
\def\essB{{S^{(1)}_{\ell}}}
\def\essC{{S^{(2)}_{\ell}}}

\begin{document}
\onecolumn
%
%%%%%%%%%%%%%%%%%%%%%%%%%%%%%%%%%%%%%%%%%%%%%%%%%%%
\title[Morphology of Weak Lensing Convergence Maps]
{Morphology of Weak Lensing Convergence Maps}
\author[Munshi et. al.]
{D. Munshi$^{a}$, T. Namikawa$^{b}$, J. D. McEwen$^{a}$, T. D. Kitching$^{a}$, F. R. Bouchet$^c$ \\
$^{a}$  Mullard Space Science Laboratory, University College London,
  Holmbury St Mary, Dorking, Surrey RH5 6NT, UK \\
$^{b}$ Department of Applied Mathematics and Theoretical Physics,
  University of Cambridge, Wilberforce Road, Cambridge CB3 OWA, UK \\
 $^{c}$ Institut d’Astrophysique de Paris, UMR 7095, CNRS \& Sorbonne Universit,
               98 bis Boulevard Arago, F-75014 Paris, France \\}
\maketitle
\begin{abstract}
  {We study the morphology of convergence maps by  perturbatively reconstructing their Minkowski Functionals (MFs).
    We present a systematics study using a set of three {\em generalised} skew-spectra 
    as a function of source redshift and smoothing angular scale.    
    Using an approach based on pseudo-$S_{\ell}$s (PSL) we show how
    these spectra will allow reconstruction of MFs in the presence
    of an arbitrary mask and inhomogeneous noise in an unbiased way. 
    Our theoretical predictions are based on a recently introduced fitting function to the bispectrum.
    We compare our results against state-of-the art numerical simulations and
    find an excellent agreement. The reconstruction can be
    carried out in a controlled manner as a function of angular harmonics $\ell$ and source redshift $z_s$ which allows
    for a greater handle on any possible sources of non-Gaussianity.
    Our method has the advantage of estimating the topology of convergence
    maps directly using shear data. We also study weak lensing convergence maps inferred from
    Cosmic Microwave Background (CMB) observations; and we
    find that, though less significant at low redshift, the {\em post-Born corrections} play an important
    role in any 
    modelling of the non-Gaussianity of convergence maps at higher redshift. 
    We also study the cross-correlations
    of estimates from different tomographic bins.}
\end{abstract}
%%%%%%%%%%%%%%%%%%%%%%%%%%%%%%%%%%%%%%%%%%%%%%%%%%%%%%%%%%%%%%%%
%
%
%%%%%%%%%%%%%%%%%%%%%%
\section{Introduction}
\label{sec:intro}
%%%%%%%%%%%%%%%%%%%%%%%
The recently completed Cosmic Microwave Background (CMB) experiments
such as the
Planck Surveyors\footnote{{\href{http://http://sci.esa.int/planck/}{\tt http://http://sci.esa.int/planck/}}}\citep{Pl13,x18} has provided us a standard model of cosmology.
However, many of the outstanding questions including, e.g., but not limited to, the nature of dark matter (DM)
and dark energy (DE) as well as possible modification of
General Relativity (GR) on cosmological scales \citep{MG1,MG2} or the exact nature of neutrino mass hierarchy \citep{nu} 
still remains unclear. 
The significant increase in precision achieved by stage-IV CMB and large scale structure surveys
will allow us to answer some of these questions.
It is expected that the ongoing weak lensing surveys 
Canada-France-Hawaii Telescope (CFHTLS\footnote{\href{http://www.cfht.hawai.edu/Sciences/CFHLS/}{\tt http://www.cfht.hawai.edu/Sciences/CFHLS}}),
Dark Energy Surveys\footnote{\href{https://www.darkenergysurvey.org/}{\tt https://www.darkenergysurvey.org/}}\citep{DES}
, Dark Energy Spectroscopic Instruments\footnote{\href{http://desi.lbl.gov}{\tt http://desi.lbl.gov}},
Prime Focus Spectrograph\footnote{\href{http://pfs.ipmu.jp}{\tt http://pfs.ipmu.jp}},
Kilo-Degree Survey (KIDS, \cite{KIDS}) and  stage-IV large scale structure (LSS)
surveys such as {\em Euclid}\footnote{\href{http://sci.esa.int/euclid/}{\tt http://sci.esa.int/euclid/}}\citep{Euclid},
Rubin Observatory\footnote{\href{http://www.lsst.org/llst home.shtml}{\tt {http://www.lsst.org/llst home.shtml}}}\citep{LSST_Tyson},
Roman Space Telescope\citep{WFIRST} will provide answers to many of the questions that cosmology is facing.
 
Weak lensing is responsible for the minute shearing and magnification in the images of the
distant galaxies by the intervening large-scale structure
allow us to extract information about clustering of the intervening  mass distribution in the Universe \citep{R19,K15,review,BS}.
Weak lensing also leaves its imprints on the observed CMB sky.
The weak lensing surveys are complementary to the galaxy surveys such as
Baryon Oscillation Spectroscopic Survey\footnote{\href{http://www.sdss3.org/surveys/boss.php}{\tt http://www.sdss3.org/surveys/boss.php}}\citep{SDSSIII},
Extended Baryon Oscillation Spectroscopic Survey\citep{eBoss}
or WiggleZ\footnote{\href{http://wigglez.swin.edu.au/}{\tt http://wigglez.swin.edu.au/}}\citep{WiggleZ} as they
provide an unbiased picture of the underlying dark matter distribution whereas the galaxies and other
tracers can only provide a biased picture \citep{bias_review}.

However, weak lensing observations are sensitive to small scales where clustering is nonlinear and non-Gaussian \citep{bernardeau_review}.
Indeed, the statistical estimates of cosmological parameters based on power spectrum analysis is typically degenerate in
cosmological parameter, e.g., $\sigma_8$ and $\Omega_{\rm M}$. External data sets, e.g., CMB as well as
tomographic or 3D \citep{3D} information is typically used to lift the degeneracy. However, an alternative procedure
would be to use high-order statistics of observables that probe the nonlinear regime\citep{higher1,higher2,higher3}.
Even in the absence of any primordial non-Gaussianity,
the gravitational clustering induces mode coupling that results in a secondery non-Gaussianity which
is more pronounced at the smaller scales where weak lensing surveys are sensitive. Thus a
considerable amount of effort has been invested in understanding the gravity induced secondary non-Gaussianity
from weak lensing surveys. These statistics include the lower order cumulants \citep{MunshiJain1} and their correlators \citep{MunshiBias};
the multispectra including the skew-spectrum \citep{MunshiHeavens} and kurtosis spectra \citep{AlanTri}
as well as the entire PDF \citep{MunshiJain2} and
the statistics of hot and cold spots.
The future surveys such as the {\em Euclid} survey will be particularly interesting in this regard.
With its large fraction of sky-coverage it will be able to detect the gravity induced non-Gaussianity with a very
high signal-to-noise (S/N).  It is also worth mentioning here that, in addition to breaking the degeneracy
in cosmological parameters the higher-order statistics is also
important in understanding the covariance of lower-order estimators. \citep{MunshiBarber1,MunshiBarber2,MunshiBarber3,MunshiBarber4}

Topological estimators such as the Minkowski Functionals (MFs)
are also important diagnostics in this direction as they carry information at all-order.
The MFs have been extensively developed as a statistical tool in a cosmological setting for both
2-dimensional (projected) and 3-dimensional (redshift) surveys. The MFs have analytically known results for
a Gaussian random field making them suitable for studies of non-Gaussianity. Examples of such
studies include CMB data (\cite{Natoli10,HikageM08,Novikov00,Schmalzing98, Bouchet13, PlanckNG, PlanckNG2}),
large scale structure
(\cite{Gott86,Coles88,Gott89,Melott89,Moore92,Gott92,Canavezes98,Schmalzing00,Kerscher01,Park05,Hikage08,Hikage06,Hikage02}),
weak lensing (\cite{Matsubara01,Sato01,Taruya02,Waerbeke}),
Sunyaev-Zel'dovich (SZ) maps \citep{MunshiSZ11},
21cm (\cite{Gleser06}) and N-body simulations (\cite{Schmalzing00,Kerscher01}).
Note that this is an incomplete list of references and we have selected a sample of
representative papers from the literature. The MFs are spatially defined topological statistics and,
by definition, contain statistical information of all orders. This makes them complementary to the
polyspectra methods that are defined in Fourier space. It is also possible that the two approaches
will be sensitive to different aspects of non-Gaussianity and systematic effects although in
the weakly non-Gaussian limit it has been shown that the MFs reduce to a weighted probe of the bispectrum (\cite{Hikage06}).
In addition to providing cosmological information, MFs can also be useful diagonistics of any unknown
systematics as well as baryonic contamination which are expected to affect weak lensing observables \citep{Waerbeke2}.

This paper is organised as follows.
The Minkowski Functionals are reviewed in \textsection\ref{sec:Mink}.
Our notations for the weak lensing statistics in projection are described in \textsection\ref{sec:power_bispec}.
The generalised skew-spectra are expressed in terms of the bispectrum in \textsection\ref{sec:gen_skew}.
The fitting function we use for our reconstruction is described in \textsection\ref{sec:fit}.
A very brief description of the simulations is provided in \textsection\ref{sec:simu}.
We discuss the results in \textsection\ref{sec:disc}.
The conclusions are presented in \textsection\ref{sec:conclu}.
%
%
%%%%%%%%%%%%%%%%%%%%%%%%%%%%%%%
\section{Minkowski Functionals}
\label{sec:Mink}
%%%%%%%%%%%%%%%%%%%%%%%%%%%%%%%
%
The MFs are related to Hadwiger's theorem \cite{Hadwiger59} in integral geometry
framework which asserts that a set of $d+1$ functionals
can provide all necessary information of a random field in $d$-dimensional space. These functionals
are a unique set of morphological estimators that are {\em motion-invariant}
and obey properties such as {\em convex-continuity} as well as {\em additivity}.
These properties are important for computing morphological estimators
from a pixelized map. The MFs are defined over an excursion set $\Sigma$
for a given threshold $\nu$ and are expressed in terms of weighted curvature integrals.

In two dimension (2D) the three MFs are defined and
can be expressed using the following notations
of \cite{Hk08}:

\ben
&& V_0(\nu) = \int_{\Sigma} da; \quad V_1(\nu) = {1 \over 4}\int_{\partial\Sigma} dl;
\quad V_2(\nu) = {1 \over 2\pi}\int_{\partial \Sigma} {\cal K} dl.
\een

\n Following the standard notation in cosmological literature, we use  $da$, $dl$ to denote the surface area and line elements for an
excursion set $\Sigma$ and its boundary $\partial \Sigma$
respectively that crosses a threshold. The MFs $V_k(\nu)$ correspond to the area of the
excursion set $\Sigma$, the length of its boundary $\partial\Sigma$
as well as the integral of curvature $\cal K$ along its boundary
which is also related to the genus $g$ and hence the Euler
characteristics $\chi$.

The Minkowski Functionals can be employed to quantify deviations from
Gaussianity. At leading order the MFs can be
constructed completely from the knowledge of the bispectrum alone.

The behaviour of the MFs for a random Gaussian field is well known and is given by
Tomita's formula \citep{Tom86}. The MFs are denoted by $V_k(\nu) (k= 0,1,2)$ 
for a threshold $\nu= \myf/\sigma_0$, where $\sigma^2_0=\la \myf^2
\ra$ can be decomposed into two different contributions, Gaussian
$V^G_k(\nu)$ and perturbative non-Gaussian contribution $\delta V_k(\nu)$:
\ben
&& V_k(\nu)= V_k^G(\nu) + \delta V_k(\nu).
\een
We are primarily
interested in the gravity induced non-Gaussian contribution, i.e. $\delta V_k(\nu)$ \citep{Hk08},
\ben
&& V^G_k(\nu) = A \exp \left ( -{\nu^2 \over 2}\right )  H_{k-1}(\nu); \quad\quad\\
&& \delta V_k(\nu) = A \exp \left ( -{\nu^2 \over 2}\right )
\left [ \delta V_k^{(2)}(\nu)\sigma_0 + \delta V_k^{(3)}(\nu)\sigma_0^2 + \delta V_k^{(4)}(\nu)\sigma_0^3 + \cdots \right ].
\label{eq:edge}
\een
where $H_k(\nu)$ is the Hermite polynomials. Following the notations introduced in \cite{Hk08} we have separated out
a normalisation factor $A$ in these
expressions which is given by the generalised variance parameter $\sigma^2_0$ and $\sigma_1^2$:
\ben
&& A = {1 \over (2\pi)^{(k+1)/2}} {\omega_2 \over \omega_{2-k}\omega_k}\left ( \sigma_1 \over \sqrt 2 \sigma_0 \right )^k.
\label{eq:v_k}
\een
Here, $\omega_k = \pi^{k/2}/\Gamma({k/2}+1)$ is the volume of a $k$-dimensional unit ball.
For projected weak lensing convergence maps in 2D we only need $\omega_0=1$, $\omega_1=2$ and $\omega_2=\pi$.
The coefficient depend only on the power spectrum of the perturbation
through $\sigma_0$ and $\sigma_1$. These quantites are defined through the following expression:
\ben
&& \sigma_j^2 = {1\over 2\pi} \sum_{\ell} [\ell(\ell+1)]^j (2l+1) {\cal C}_{\ell} W^2_{\ell}.
\label{eq:define_variance}
\een
Here ${\cal C}_{\ell}$ is the angular power spectrum of the underlying field and $W_{\ell}$ is the window function used to smooth a map.
A more through discussion will be presenetd in the follwoing section for $\kappa$ maps.
At the level of the bispectrum the perturbative corrections are determined by three generalised skewness paramters $S^{(k)}$ \citep{Hk08}:
\ben
&& \delta V_k^{(2)}(\nu) = \left [ \left \{  {1 \over 6} S^{(0)} H_{k+2}(\nu) + {k \over 3} S^{(1)} H_k(\nu) +
  {k(k-1) \over 6} S^{(2)} H_{k-2}(\nu)\right \} \right ];
%%\quad\quad\\
%%&& A = {1 \over (2\pi)^{(k+1)/2}} {\omega_2 \over \omega_{2-k}\omega_k}\left ( \sigma_1 \over \sqrt 2 \sigma_0 \right )^k.
\label{eq:v_k}
\een
The skewness parameters can also be expressed as \citep{Waerbeke}:
\ben
&& S^{(0)} = {\la \kappa^3 \ra\over \sigma_0^2}; \quad S^{(1)} = {\langle\kappa^2 \nabla^2 \kappa \rangle\over \sigma_0^2\sigma_1^2};
\quad S^{(2)} \equiv {\langle |\nabla \kappa |^2 \nabla^2 \kappa \rangle \over \sigma_1^2}.
\label{eq:define_skewness}
\een
Here, $S^{(0)}$ is the ordinary skewness parameter where as $S^{(1)}$ and $S^{(2)}$ are its higher-order generalisations.
At next order a set of four kurtosis parameters can be used to expressed the next-order correlations \citep{Waerbeke}.
The primary motivation of this article is to reconstruct these generalised skewness parameters using
spectra associated with them that allows to estimate them from surveys in the presence of
complicated mask and noise. We will borrow the analytical tools developed in \citep{Waerbeke}.
%
%
%%%%%%%%%%%%%%%%%%%%%%%%%%%%%%%%%%%%%%%%%%%%%%%%%%%%
\section{Weak Lensing Power spectrum and Bispectrum}
\label{sec:power_bispec}
%%%%%%%%%%%%%%%%%%%%%%%%%%%%%%%%%%%%%%%%%%%%%%%%%%%
%
The weak lensing convergence denoted as $\kappa$ can be expressed in terms of a line-of-sight (los) integration
of three-dimensional (3D) density contrast $\delta$
\ben
&& \kappa(\bm\theta, r_s) = \int_o^{r_s}\, dr\, \omega(r,r_s) \delta(\bm\theta, r); \quad
\omega(r,r_s) = {3\over 2 \,a}{H_0^2 \over c^2}\Omega_M {d_A(r-r_s)\over d_A(r)d_A(r_s)};
\een
In our notation $r=|{\bm r}|$ denotes the {\em comoving} radial distance to the source and $\bm\theta$ denotes the angular position on the sky,
The background cosmology is specified in terms of  $\Omega_{\rm M}$ which denotes the cosmological matter density parameter
(that describes the total matter density in units of the critical density), 
$H_0$ which denotes the Hubble constant; $c$ is the speed of light, and $a = 1/(1+z)$ denotes the scale factor at a redshift $z$.
The comoving angular diameter distance at a comoving radial  distance $r$ is represented as $d_A(r)$.
The source plane is assumed to be at a redshift $z_s$, or equivalently at comoving radial distance $r_s$.
To simplify the analysis we will ignore source distribution and photometric redshift errors. We will
focus on the morphological estimators as a function as a function of $z_s$.

For the smoothed convergence $\kappa$, the mean is zero, $\langle\kappa(\bm\theta) \rangle=0$, and
using a spherical harmonic decomposition of $\kappa(\bm\theta)$, using spherical harmonics $Y_{\ell m}(\bm\theta)$ as the basis functions,
$\myf(\bm\theta) = \sum_{\ell m} \myf_{\ell m} Y_{\ell m}(\bm\theta)$,
we can define its angular power spectrum $\myC_l$ in terms of the harmonic coefficients $\kappa_{\ell m}$
$\la \myf_{\ell m} \myf^*_{\ell'm'} \ra = \myC_\ell\delta_{\ell\ell'}\delta_{mm'}$
which is a sufficient statistical characterization of a Gaussian field. 
\ben
&& {\cal C}_{\ell} = \int_0^{r_s} dr {w^2(r,r_s) \over d_A^2(r)} P_{} \left ({\ell \over d_A(r)}; r \right ).
\een
The convergence bispectrum ${\cal B}$ can likewise be expressed using the following los integration of the bispectrum
of the density contrast $\delta$ denoted as $B_{\delta}$ (see \citep{review}): 
\ben
&& \la \kappa_{\ell_1 m_1}\kappa_{\ell_2 m_2}\kappa_{\ell_3m_3}\ra_c \equiv 
{\cal B}_{\ell_1\ell_2\ell_3}
\left ( \begin{array} { c c c }
     \ell_1 & \ell_2 & \ell_3 \\
     m_1 & m_2 & m_3  
\end{array} \right ).
\een
The matrix above represents a  Wigner $3\rm j$ symbol and the angular brackets here represent ensemble averaging.
The angular brackets represent ensemble averaging. This particular form is employed as it 
preserves the the rotational invariance of the three-point correlation function.

 The  Wigner $3j$-symbol, which is nonzero only when the triplets $(\ell_1, \ell_2, \ell_3)$ satisfy the
 {\em triangularity} condition $|\ell_1 -\ell_2| \le \ell_3 \le \ell_1 + \ell_2$ as well as the condition that the sum $\ell_1 + \ell_2 + \ell_3$ is even.
 This ensures the parity invariance of the bispectrum and neglect presence of any parity violating physics.
This selection rule is imposed by the invariance of the field under spatial inversion.
 Indeed, the parity violating contributions at the level of the  bispectrum can be obtained by
including both the (so-called) Electric $(\rm E)$ and Magnetic $(\rm B)$ modes \citep{higher1}. This can be used to detect any possible parity violating physics as well as other systematic effects.

The convergence bispectrum ${\cal B}^{}$ is expressed in terms of the bispectrum for the density contrast: $B_{}$: 
\ben
&& {\cal B}_{\ell_1\ell_2\ell_3} = I_{\ell_1\ell_2\ell_3}\int_0^{r_s} dr {w^3(r,r_s) \over d_A^4(r)}
B_{}\left ({\ell_1 \over d_A(r)}, {\ell_2 \over d_A(r)}, {\ell_3 \over d_A(r)} ; r \right ) \\
&& I_{\ell_1\ell_2\ell_3} = \sqrt{(2\ell_1+1)(2\ell_2+1)(2\ell_3+1) \over 4\pi}\left ( \begin{array} { c c c }
     \ell_1 & \ell_2 & \ell_3 \\
     0 & 0 & 0
\end{array} \right ).
\een
The cross-spectrum ${\cal C}^{\alpha\beta}_{\ell}$ and mixed bispectrum ${\cal B}^{\alpha\beta}_{\ell_1\ell_2\ell_3}$ involving two
topographic bins $\alpha$ and $\beta$ have the following form:
\bes
\ben
&& {\cal C}^{\alpha\beta}_{\ell} = \int_0^{r_{\rm min}} dr {\omega_\alpha(r)\omega_\beta(r) \over d_A^2(r)}
P\left ({{l} \over d_A(r)}; r \right );\\
&& {\cal B}^{\alpha\beta}_{\ell_1\ell_2\ell_3} = I_{\ell_1\ell_2\ell_3} \int_0^{r_{\rm min}} dr {\omega_{\alpha}^1(r)\omega_{\beta}^2(r) \over d_A^4(r)}
{B}_{}\left ({{\ell}_{1} \over d_A(r)},{{\ell}_{2} \over d_A(r),},
{{\ell}_{3} \over d_A(r)}; r \right); \quad r_{min} = min(r_{\alpha},r_{\beta}); \label{eq:tomo_skew_a}\\
&& w_i(r) := {3 \Omega_{\rm M} \over 2} {H_0^2 \over c^2} a^{-1} {d_A(r) d_{A}({r_{si}-r}) \over d_A(r_{si})}; \quad i\in\{\alpha,\beta\}.
\label{eq:tomo_skew_b}
\een
\ees
Using these expression we will next construct the generalised skew-spectra that are useful in
constructing the MFs.
%
%
%%%%%%%%%%%%%%%%%%%%%%%%%%%%%%%%%%
\section{Generalised Skew-Spectra}
\label{sec:gen_skew}
%%%%%%%%%%%%%%%%%%%%%%%%%%%%%%%%%%
%
Individual triplets of harmonics $(\ell_1,\ell_2,\ell_3)$ defines a triangle in the harmonic domain
and specify a bispectral mode. The skew-spectra defined below are summed over all
possible configuration of the bispectrum by keeping one side of the triangle fixed.
Following \citep{Waerbeke} we introduce the generalised skew-spectra $S_\ell^{(i)}$:
\bes
\ben
&& S_\ell^{(0)} =
{1 \over 12 \pi \sigma_0^4} {1\over 2\ell+1}\sum_m {\rm Real}\{ [\kappa^2]_{\ell m}[\kappa]^*_{\ell m} \}
 ={1 \over 12 \pi \sigma_0^4} \sum_{\ell_1\ell_2} \myB_{\ell\ell_1\ell_2}J_{\ell\ell_1\ell_2}W_{\ell}W_{\ell_1}W_{\ell_2};  \label{eq:s0}\\
 && S_\ell^{(1)} = {1 \over 16 \pi \sigma_0^2\sigma_1^2} {1\over 2\ell+1}\sum_m {\rm Real}\{ [\kappa^2]_{\ell m}\nabla^2[\kappa]^*_{\ell m} \}
 = {1 \over 16 \pi \sigma_0^2\sigma_1^2}  \ell(\ell+1)\sum_{\ell_i}  \myB_{\ell\ell_1\ell_2}J_{\ell\ell_1\ell_2}
W_{\ell}W_{\ell_1}W_{\ell_2};   \label{eq:s1}\\
&& S_\ell^{(2)}=  {1 \over 8 \pi \sigma_1^4} {1\over 2\ell+1}\sum_m {\rm Real}\{ [\nabla\kappa\cdot\nabla\kappa]_{\ell m}[\kappa^2]^*_{\ell m} \}\nn \\
&& \quad\quad  = {1 \over 8 \pi \sigma_1^4} \sum_{\ell_i}
\Big [ [\ell(\ell+1)+\ell_1(\ell_1+1) - \ell_2(\ell_2+1)]\ell_2(\ell_2+1) \Big ] 
\myB_{\ell\ell_1\ell_2}J_{\ell\ell_1\ell_2}W_{\ell}W_{\ell_1}W_{\ell_2}. \label{eq:s2}
\een
We have introduced the following notations above:
\ben
&& J_{\ell_1\ell_2\ell_3} \equiv {I_{\ell_1\ell_2\ell_3} \over 2\ell_3+1} = \sqrt{(2\ell_1+1)(2\ell_2+1) \over (2\ell_3+1) 4 \pi }\left ( \begin{array}{ c c c }
     \ell_1 & \ell_2 & \ell_3 \\
     0 & 0 & 0
\end{array} \right); \\
&& W_{\ell} = \exp\left [-{}\ell(\ell+1){\theta_{s}^2 \over 8\ln 2 }\right ].
\een
\ees
We will study these spectra using numerical simulations and test them against theoretical predictions that rely on a fitting function
based approach. We will use a Gaussian window function $ W_{\ell}$ in our study but the expressions are valid for arbitrary window function,
including the tophat or compensated window (filter) functions. The one-point skewness parameters $S^{(i)}$ can be recovered
from their respective skew-spectra, which were used in Eq.(\ref{eq:define_skewness}):
\ben
&& S^{(i)} = {1 \over 4\pi}\sum_{l}(2\ell+1)S^{(i)}_\ell. \quad
%\sigma_j^2 = {1 \over 4\pi}\sum_l (2l+1)[l(l+1)]^j \myC_l W_l^2
\label{skew_spectra}
\een
Expressions for the skew-spectra in  Eq.(\ref{eq:s0})-Eq.(\ref{eq:s2}) can also be generalised to include cases where instead of
individual bins two different bins are cross-correlated.
\ben
&& S_\ell^{(0)\alpha\beta} =
{1\over 2\ell+1}\sum_m {\rm Real}\{ [\kappa_\alpha^2]_{\ell m}[\kappa_\beta]^{*}_{\ell m} \}
={1 \over 12 \pi \sigma_0^4} \sum_{\ell_1\ell_2} \myB^{\alpha\beta}_{\ell\ell_1\ell_2}J_{\ell\ell_1\ell_2}W_{\ell}W_{\ell_1}W_{\ell_2}  \label{eq:s0_cross}.
\een
Similar expressions can be obtained for the other skew-spectra by replacing $\myB_{\ell\ell_1\ell_2}$ by
$\myB^{\alpha\beta}_{\ell\ell_1\ell_2}$ in Eq.(\ref{eq:s1})-Eq.(\ref{eq:s2}).
The mixed bispectra $\myB^{\alpha\beta}_{\ell\ell_1\ell_2}$ is defined in Eq.(\ref{eq:tomo_skew_a}).
Notice that by construction  $S_\ell^{(i)\alpha\beta} \ne S_\ell^{(i)\beta\alpha}$ as $\myB^{\alpha\beta}_{\ell\ell_1\ell_2} \ne \myB^{\beta\alpha}_{\ell\ell_1\ell_2} $.

Although we have adopted an harmonic approach, equivalent information about the non-Gaussianity can also be
obtained by studying the corresponding {\em collapsed} three-point correlation functions.
This approach will be more efficient for surveys with smaller sky-coverage  and in the presence of a non-trivial mask: 
\ben
&& S^{(0)\alpha\beta}_{12}(\theta_{}) = \langle \kappa_{\alpha}^2({\bm\theta_1})\kappa_{\beta}(\bm\theta_2)\rangle; \quad
S^{(1)\alpha\beta}_{12}(\theta_{}) = \langle \kappa_{\alpha}^2({\bm\theta_1})\nabla^2\kappa_{\beta}(\bm\theta_2)\rangle; \quad
S^{(2)\alpha\beta}_{12}(\theta_{}) =
\langle \nabla^2\kappa_{\alpha}({\bm\theta_1})[\nabla\kappa_\beta(\bm\theta_2)\cdot\nabla\kappa_\beta(\bm\theta_2)]\rangle; \quad
\een
Due to the isotropy and homogeneity of the background Universe these correlations functions are only function
of the separation angle $\theta_{} = |\bm\theta_1 - \bm\theta_2|$. These two-point correlations
can be constructed by cross-correlating derived maps from different topographic bins  $\kappa_\alpha^2(\bm\theta)$,
$\nabla^2\kappa_\alpha(\bm\theta)$ and $\nabla\kappa_\alpha(\bm\theta)\cdot\nabla\kappa_\alpha(\bm\theta)$.
In terms of the skew spectra these correlations functions can be expressed as: 
\ben
&& S^{(i)\alpha\beta}_{12}(\theta)  = {1 \over 4\pi} \sum_{\ell} (2\ell+1) P_{\ell}(\cos \theta_{}) S^{(i)}_{\ell} \,; \quad\quad  i \in \{0,1,2 \}.
\een
Here $P_{\ell}$ denotes the Legendre polynomial of order $\ell$.
%
%%%%%%%%%% Figure -3 %%%%%%%%%%%%%%%%%%%%%%%%%
\begin{figure}
  \centering
  \includegraphics[width=0.7\textwidth]{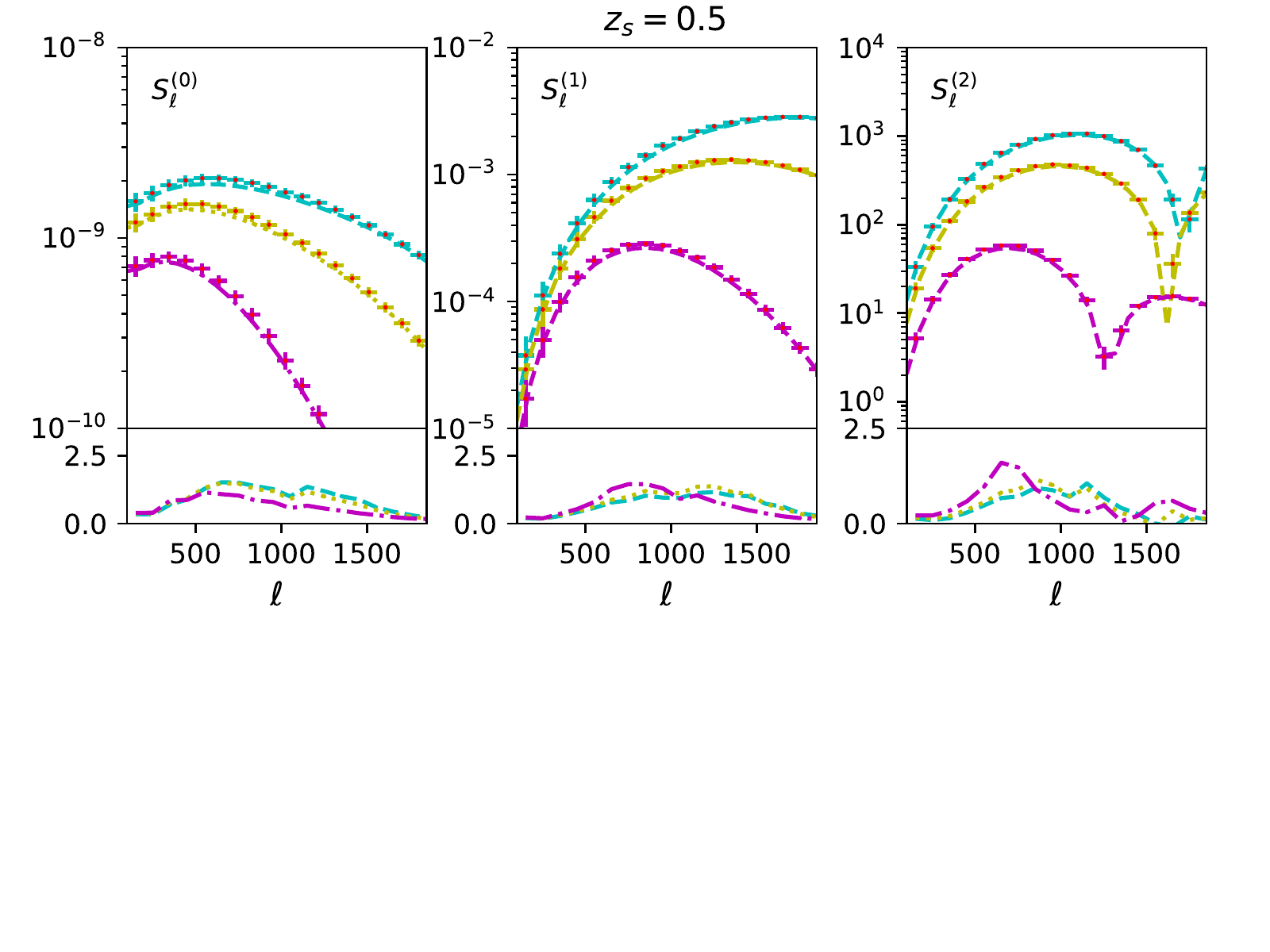}
  \vspace{-3.25cm}
  \hspace{1.cm}
  \caption{From left to right different panels depict the skew-spectra $S^{0}_{\ell}$, $S^{(1)}_{\ell}$ and $S^{(1)}_{\ell}$ respectively
    as a function of $\ell$. The data points with erro-bars in each panel are the bin-averaged values of the
    respective skew-spectra estimated from simulations.
    Different curves in each panel correspond to different smoothing angular scales.
    These generalised skew-spectra are defined in Eq.(\ref{eq:s0})-Eq.(\ref{eq:s2}). In each panel, three different
    smoothing angular scales $\theta_s= 2', 5'$ and $10'$ (from top to bottom) are shown. The source redshift is fixed at $z_s=0.5$.
    A Gaussian smoothing window was used. See text for more details. The bottom subpanels for each panel show the deviation $\Delta_l$ of simulations $S^{\rm sim}_\ell$
    from theoretical prediction $S^{\rm th}_\ell$ in units of standard deviation $\sigma_{\ell}$ computed for individual beans
    i.e. $\Delta_{\ell}=(S^{\rm th}_{\ell}-S^{\rm sim}_{\ell})/\sigma_{\ell}$. No noise or mask were used.}
  \label{fig:save_mink1}
\end{figure} 
%%%%%%%%%%%%%%%%%%%%%%%%%%%%%%%%%%%%%%%%%%%%%%%
%%%%%%%%%% Figure -3 %%%%%%%%%%%%%%%%%%%%%%%%%
\begin{figure}
  \centering
  \includegraphics[width=0.7\textwidth]{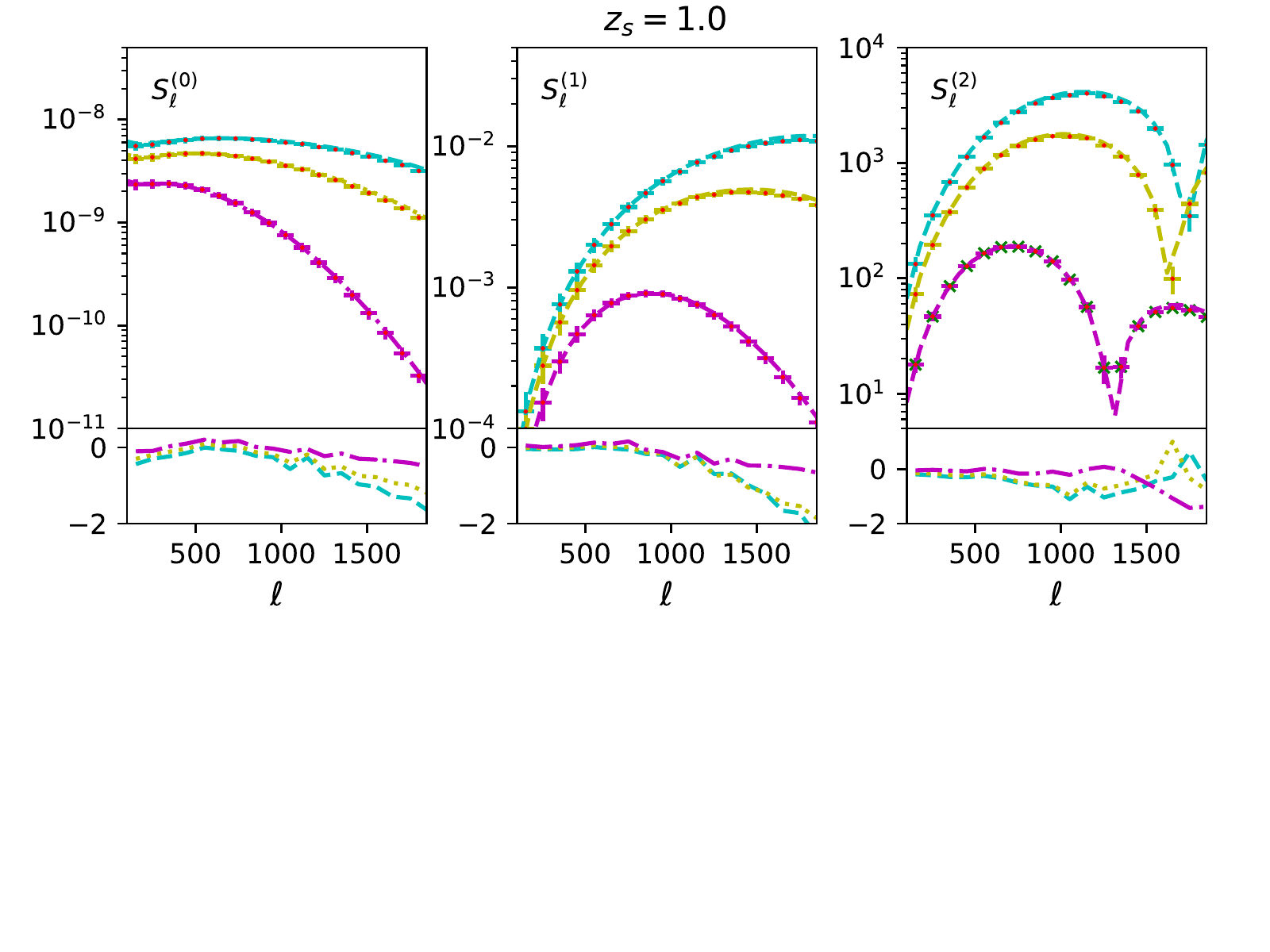}
  \vspace{-3.25cm}
  \hspace{1.cm}
  \caption{Same as Figure-\ref{fig:save_mink1}, but for $z_s=1.0$.}
  \label{fig:save_mink2}
\end{figure} 
%%%%%%%%%%%%%%%%%%%%%%%%%%%%%%%%%%%%%%%%%%%%%%%

%%%%%%%%%% Figure -3 %%%%%%%%%%%%%%%%%%%%%%%%%
\begin{figure}
  \centering
  \includegraphics[width=0.7\textwidth]{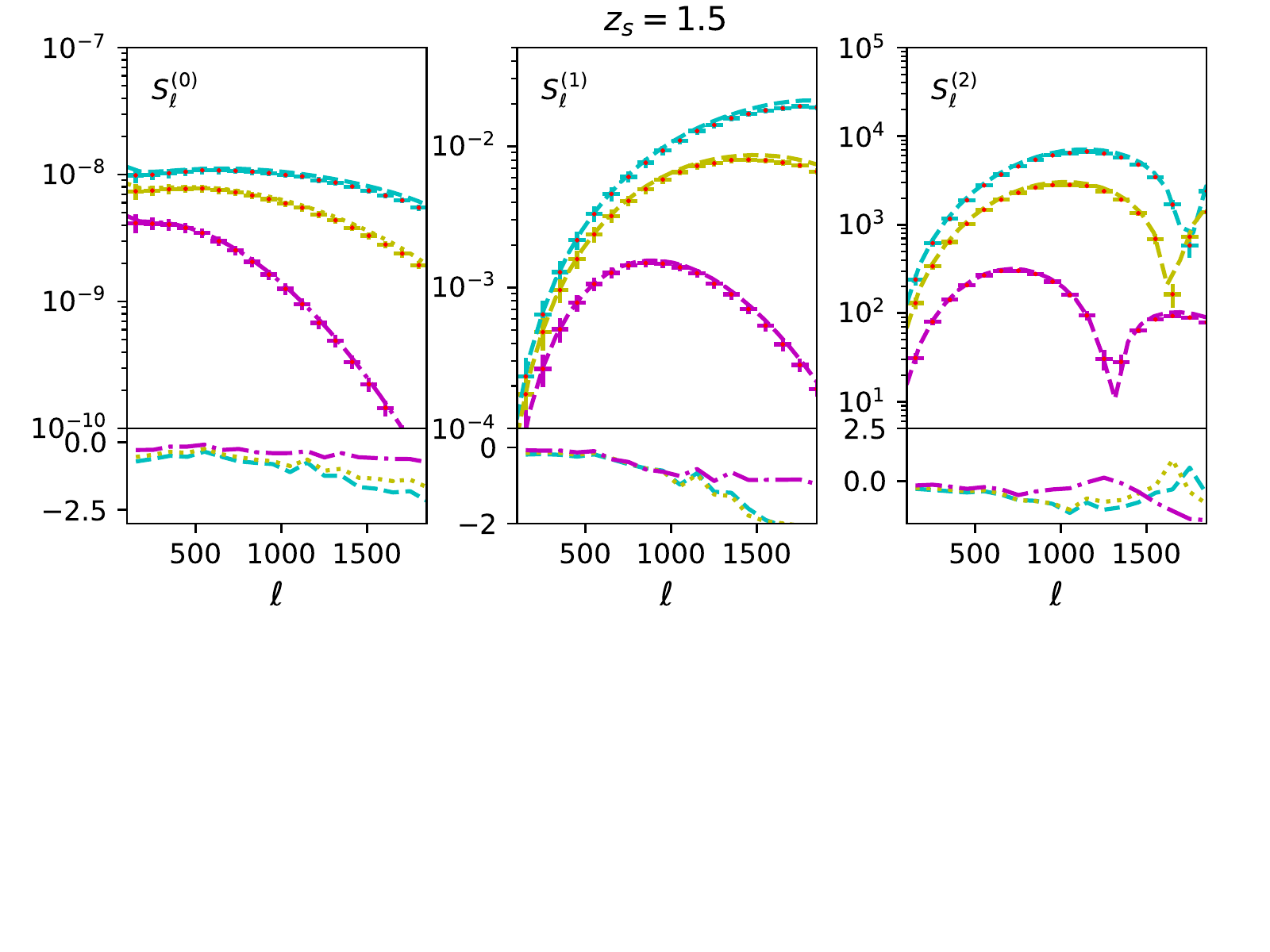}
  \vspace{-3.25cm}
  \hspace{1.cm}
  \caption{Same as Figure-\ref{fig:save_mink1}, but for $z_s=1.5$.}
  \label{fig:save_mink3}
\end{figure} 
%%%%%%%%%%%%%%%%%%%%%%%%%%%%%%%%%%%%%%%%%%%%%%%
%
%%%%%%%%%% Figure -3 %%%%%%%%%%%%%%%%%%%%%%%%%%%%
\begin{figure}
  \centering
  \includegraphics[width=0.7\textwidth]{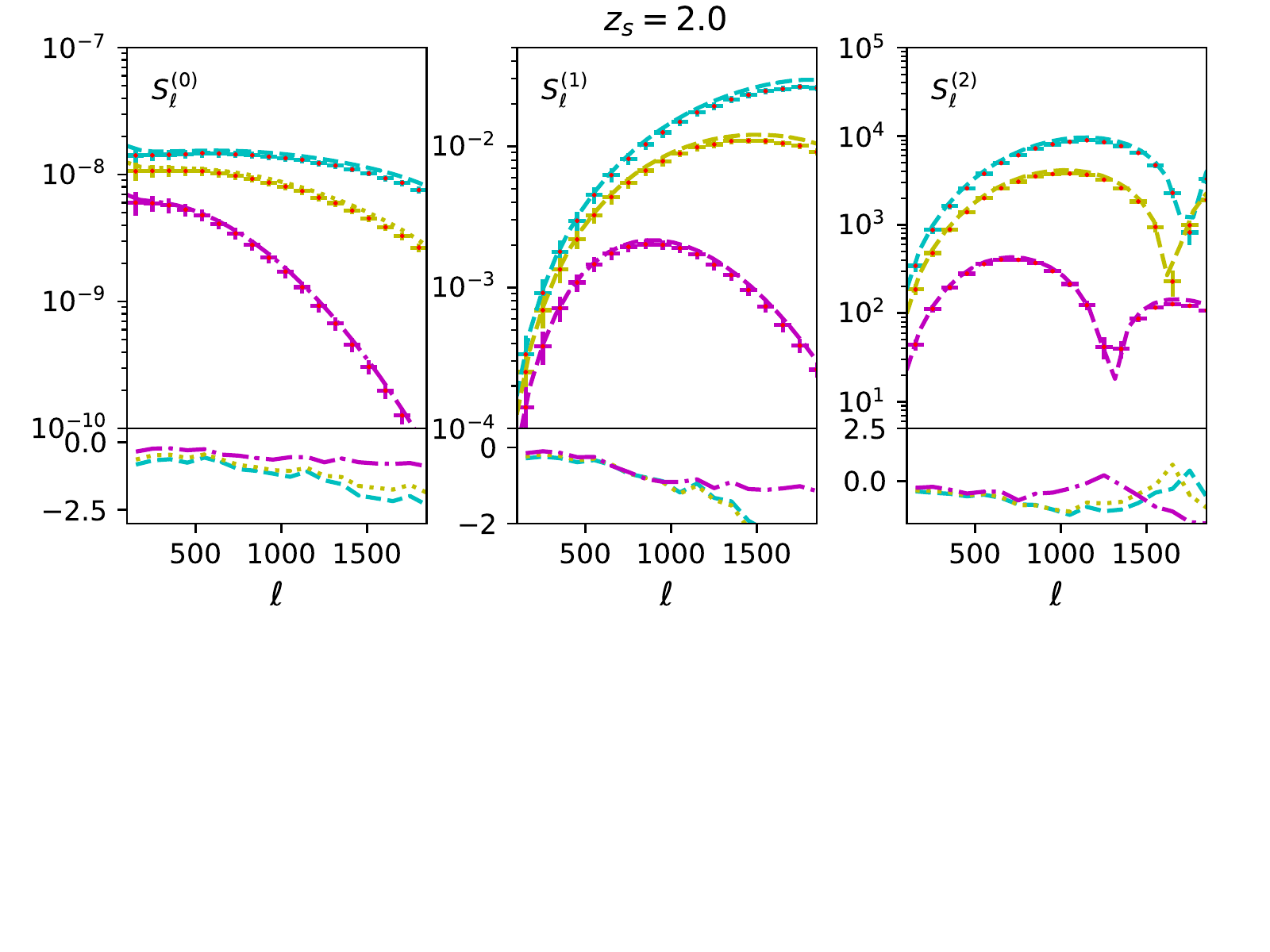}
    %\end{minipage}
  %\end{center}
  \vspace{-3.25cm}
  \hspace{1.cm}
  \caption{Same as Figure-\ref{fig:save_mink1}, but for $z_s=2.0$.}
  \label{fig:save_mink4}
\end{figure} 
%%%%%%%%%%%%%%%%%%%%%%%%%%%%%%%%%%%%%%%%%%%%%%%%%
%
So far we have assumed a full-sky coverage for estimation of the generalised skew-spectra.
However, most surveys will have a partial sky-covergage. The pseudo skew-spectrum (PSL) technique presented in \citep{BouchetSkew1}
is also valid for the generalised skew-spectra. An unbiased all-sky estimate ${\hat S}_{\ell}$ can be
constructed from the masked skew-spectra ${\tilde S}_{\ell}$ using the expression below:
\bes
\ben
&& {\tilde S}^{(i)}_{\ell} = M_{\ell\ell^{\prime}} S^{(i)}_{\ell} \label{eq:tilde}; \quad
{\hat S}^{(i)}_{\ell} = M^{-1}_{\ell\ell^{\prime}}{\tilde S}^{(i)}_{\ell} \label{eq:hat}; \quad \langle {\hat S}^{(i)}_{\ell}\rangle  = S^{(i)}_{\ell}
%\label{eq:PCL2}
\label{eq:PCL1}
\een
\ees
where the mode-coupling (mixing) matrix is given by:
\ben
&& M_{\ell\ell'} = (2\ell'+1)\sum_{\ell''}
\left ( \begin{array}{ c c c }
     \ell & \ell' & \ell'' \\
     0 & 0 & 0
  \end{array} \right)^2
{ (2\ell'' +1 )\over 4\pi} |w_{\ell''}^2|; \label{eq:MLL1}
%&& \hat{\cal S}_{\ell}^{(21)} = \sum_{\ell'}M_{\ell\ell'}^{-1} \tilde{\cal S}_{\ell'}^{(21)} \label{eq:MLL2}.
\een
Here we have introduced the power spectrum of the mask $w(\bm\theta)$, i.e., $w_{\ell} = {1/(2\ell+1)} \sum_m |w_{\ell m}|^2 $, 
constructed from the harmonic-coefficient $w_{\ell m}$ and its complex conjugate $w^*_{\ell m}$ (see \cite{BouchetSkew1}
for more detailed discussion)
Notice that a (inhomogeneous) Gaussian noise do not contribute to the generalised skew-specra though it will increase the scatter.
This PSL method will be essential for constructing morphology of weak lensing $\kappa$ maps in the
presence of a mask with non-trivial topology. 
%
%%%%%%%%%%%%%%%%%%%%%%%%%%%%%%%%%%%%%%%%%
\section{Fitting Function for Bispectrum}
\label{sec:fit}
%%%%%%%%%%%%%%%%%%%%%%%%%%%%%%%%%%%%%%%%%%
%
%
In second-order Eulerian perturbation theory the matter bispectrum $B_{}(\bk_1,\bk_2,\bk_3)$
that encodes mode coupling of the 3D density contrast in the Fourier domain can be expressed as \citep{bernardeau_review}:
\bes
\beqa
&& B(\bk_1,\bk_2,\bk_3) = 2 F_2({\bf k_1}, {\bf k_2}) P_{lin}^{}(\bk_1)P_{lin}^{}(\bk_2) + {\rm cyc.perm.}.
\een
Here $F_2$ is the kernel that encapsulates the second-order mode-mode coupling and $P_{lin}^{}(\bk)$ denotes
the linear power spectrum of the density contrast $\delta$. In a fitting function approach the analytical form of the kernel $F_2$
is generalised from the quasi-linear regime to nonlinear regime by introducing three independent coefficients
$a({n_{e}},k),b({n_{e}},k)$ and $c({n_{e}},k)$ that are determined using numerical simulations.
\ben
&& F_2({\bf k_1}, {\bf k_2}) ={5 \over 7}a(n_{e},k)a(n_{e},k)+
{1 \over 2}\left ( { {\bf k}_1 \cdot {\bf k}_2 \over  k_2^2}   +{ {\bf k}_1 \cdot {\bf k}_2 \over  k_1^2} \right ) b(n_{e},k)b(n_{e},k) +
       {2 \over 7} \left ( { {\bf k}_1 \cdot {\bf k}_2 \over k_1 k_2} \right )^2 c(n_{e},k)c(n_{e},k)
\label{eq:bispec_pert}
\eeqa
\ees
Here $n_e$ is local logarithmic slope of the power spectrum at 3D wavenumber $k$.
In the quasi-linear regime these coefficients approach unity, i.e, $a=b=c=1$. In the highly nonlinear regime,
if we set $a\ne 0$ and $b=c=0$, we recover the hierarchal form for the matter bispectrum.
The idea of a fitting function was initially proposed in \citep{Scoccimarro_fit}. It interpolates 
between the perturbative and the nonlinear regimes. It has a limited validity range of $k< 3h{\rm Mpc}^{-1}$ and $z \approx 0-1$.
The functional form of this fitting function was later improved by \citep{GilMarin}
with a rather limited validity range of $k<0.4 h{\rm Mpc}^{-1}$ and $z \approx 1.5$.
The improvement was achieved by introducing additional free parameters which are
extracted from numerical simulations. The inaccuracy of this fitting function was pointed out by \cite{BouchetShapes}.
An even more accurate fitting function was recently proposed by \citep{Ryuichi}. This new fitting function
has a validity range of $k<10 h{\rm Mpc}^{-1}$ and $z \approx 1-3$.
Its higher accuracy is important for a very accurate theoretical predictions
of secondary non-Gaussianity across the range of wavelength and redshift that will be
useful for stage-IV large scale structure experiments including {\em Euclid}.
This function has already been used in \citep{BouchetSkew1}. In our study we will use
it to compute the theoretical predictions
for our morphological estimators.

For modelling of skew-spectrum related to secondary non-Gaussianity, using halo model as well
as primordial non-Gaussianity, see \citep{Waerbeke}.
%%%%%%%%%%%%%%%%%%%%%%%
\section{Simulations}
\label{sec:simu}
%%%%%%%%%%%%%%%%%%%%%%%
%
In our numerical investigations we use the all-sky weak lensing maps described in
\citep{Ryuichi}\footnote{http://cosmo.phys.hirosaki-u.ac.jp/takahasi/allsky\_raytracing/}.
These maps were generated using ray-tracing through N-body simulations using
multiple lens planes and to generate convergence $\kappa$ as well as shear $\gamma$ maps.
They do not employ the Born approximation. The post-Born corrections are known to play an important role
at higher redshifts especially for CMB lensing.
The source redshifts used were in the range $z_s= 0.05-5.30$ at an interval of $\Delta z_s = 0.05$.
We have used the maps corresponding to source redshifts of $z_s=0.5,1.0,1.5, 2.0$ in our study.
For generating lensed CMB maps numerical simulations were replaced by Gaussian realisations of density fluctuations
in the redshift range $z_s=7.1-1100.0$. The perturbations were generated using a linear matter power spectrum.
These maps were generated using different resolution
in {\tt HEALPix}\footnote{https://healpix.jpl.nasa.gov/} format\citep{Gorski} using an equal area pixelisation scheme.
The number of pixels scales as $N_{\rm pix} = 12 N^2_{\rm side}$ with the resolution parameter $N_{\rm side}$.
We will be using maps generated at a resolution $N_{\rm side}=4096$ and used maps 
at a higher resolution for various sanity checks.
In our study we will be restricting us to $\ell \le \ell_{\rm max}$ with $\ell_{\rm max} =2000$. However, the $\ell_{\rm max}$
is kept flexible in our analytical formalism and can be used to filter out any astrophysical complexities related 
baryonic feedback \citep{Weiss19}.

The cosmological parameters used are $\Omega_{\rm CDM} = 0.233$, $\Omega_b = 0.046$,
$\Omega_{\rm M} = \Omega_{\rm CDM}+\Omega_b, \Omega_{\Lambda}=1-\Omega_{\rm M}$ and $h=0.7$. For the amplitude of
density fluctuation, $\sigma_8=0.82$, and the spectral index $n_s=0.97$ is used. These maps were recently used to analyze the bispectrum
in the context of CMB lensing \citep{Namikawa18} as well in studies of lensing induced bispectrum
in low redshift \citep{BouchetSkew1,BouchetShapes}

%
% 
%%%%%%%%%%%%%%%##########%%%%%%%
\section{Results and Discussion}
\label{sec:disc}
%%%%%%%%%%%%%%%%%%%%%%##########
%
%%%%%%%%%% Figure -3 %%%%%%%%%%%%%%%%%%%%%%%%%
\begin{figure}
  \centering
  \includegraphics[width=0.75\textwidth]{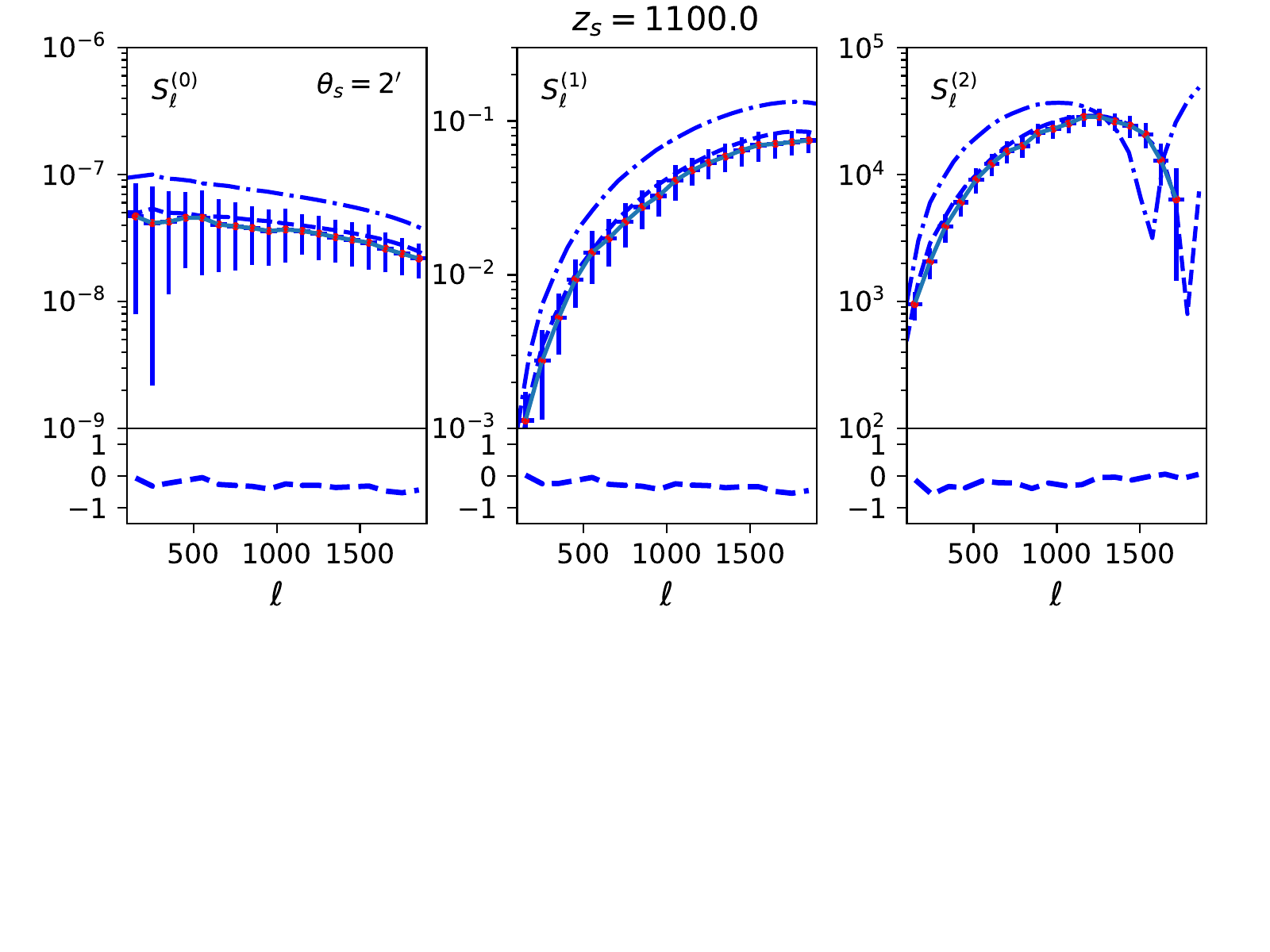}
  \vspace{-3.cm}
  \hspace{1.cm}
  \caption{The generalised skew-spectra $\kappa$ maps are shown
    for $z_s=1100.$. From left to right we show results for
    the skew-spectra $S^{(0)}_{\ell}$, $S^{(1)}_{\ell}$ and  $S^{(2)}_{\ell}$ as a function of
    $\ell$. These generalised skew-spectra are defined in Eq.(\ref{eq:s0})-Eq.(\ref{eq:s2}).
    The $\kappa$ maps are
    inferred from CMB temperature maps. The dashed and solid lines represents
    the theoretical predictions based on Born- and post-Born approximation.
    The importance of post-Born approximation is more pronounced at higher redshift.
    The smoothing angular scale is fixed at $\theta_s=2'$. Results are obtained using one all-sky map.
    No noise was included. An all-sky coverage was assumed. The error-bars were computed using the scatter
    within the bin fixed at $\delta\ell =100$.}
  \label{fig:cmb_skew1}
\end{figure}  
%%%%%%%%%%%%%%%%%%%%%%%%%%%%%%%%%%%%%%%%%%%%%%%
%
%%%%%%%%%% Figure -3 %%%%%%%%%%%%%%%%%%%%%%%%%
\begin{figure}
  \centering
  \includegraphics[width=0.7\textwidth]{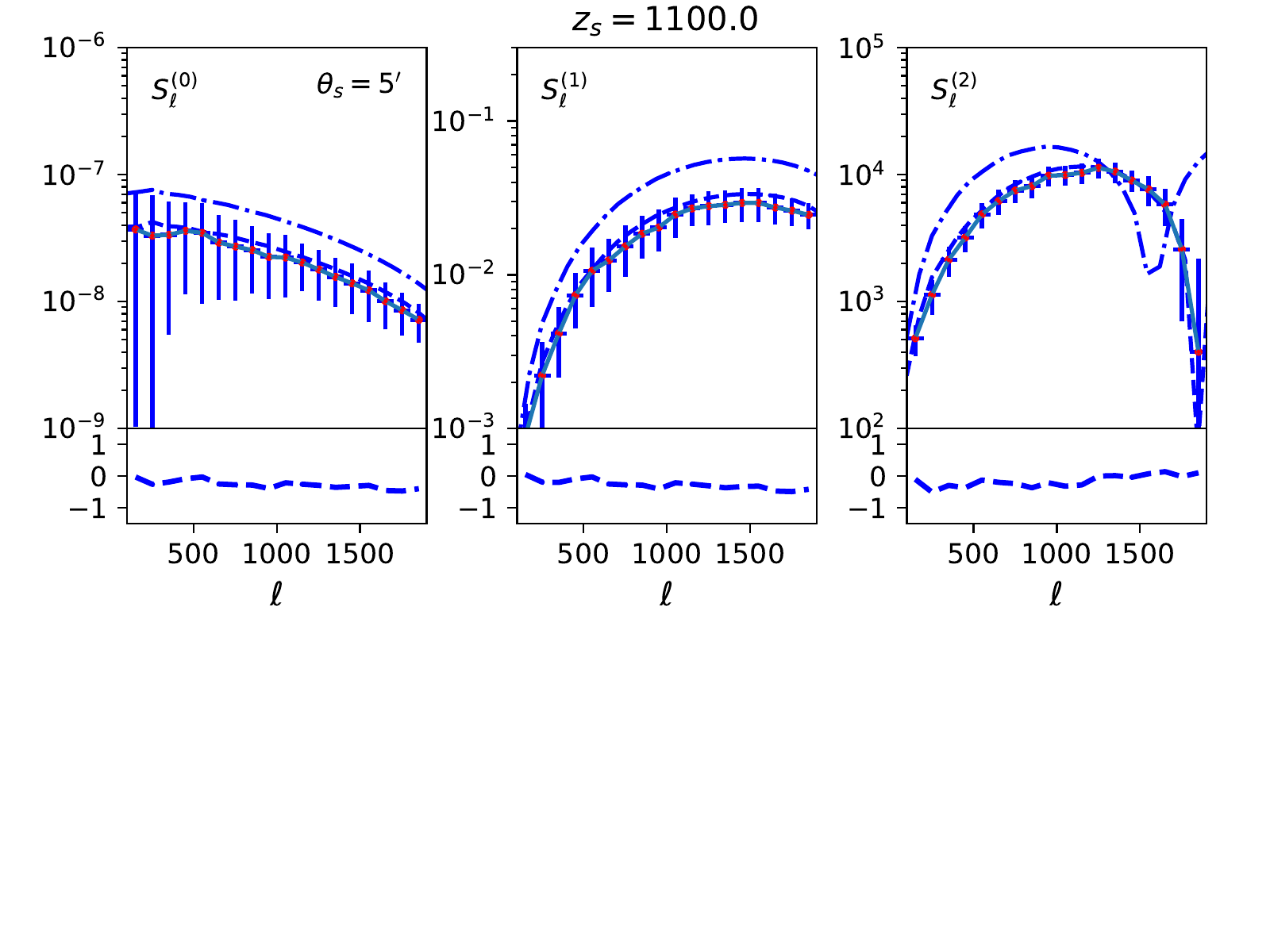}
  \vspace{-3.25cm}
  \caption{Same as Figure-\ref{fig:cmb_skew1}, but for $\theta_s=5.0'$.}
  \label{fig:cmb_skew2}
\end{figure} 
%%%%%%%%%%%%%%%%%%%%%%%%%%%%%%%%%%%%%%%%%%%%%%%

%%%%%%%%%% Figure -3 %%%%%%%%%%%%%%%%%%%%%%%%%
\begin{figure}
  \centering
  \includegraphics[width=0.7\textwidth]{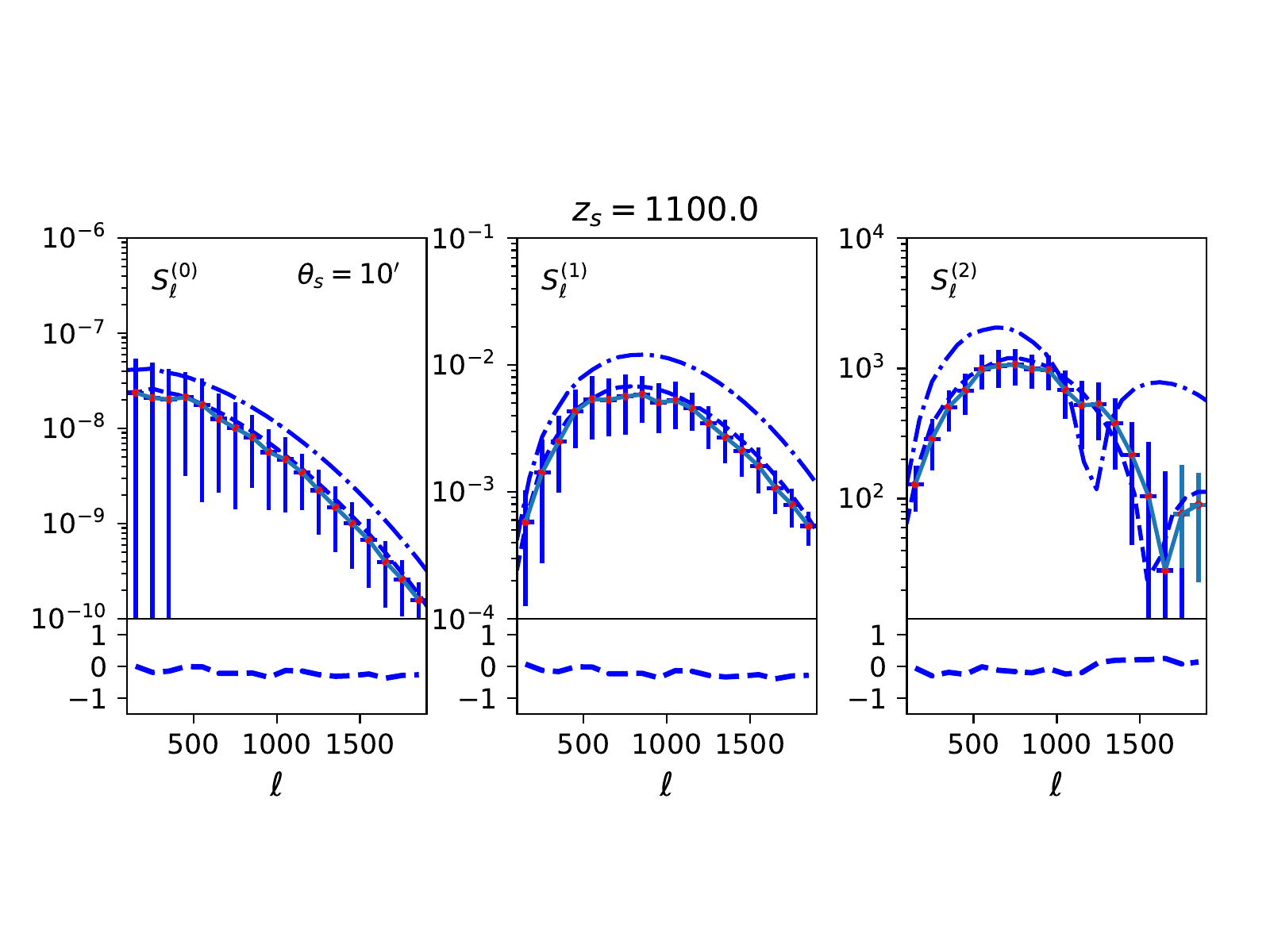}
  \vspace{-1.25cm}
  \caption{Same as Figure-\ref{fig:cmb_skew1}, but for $\theta_s=10.0'$.}
  \label{fig:cmb_skew3}
\end{figure} 
%%%%%%%%%%%%%%%%%%%%%%%%%%%%%%%%%%%%%%%%%%%%%
 
\begin{figure}
  \centering
  \includegraphics[width=0.7\textwidth]{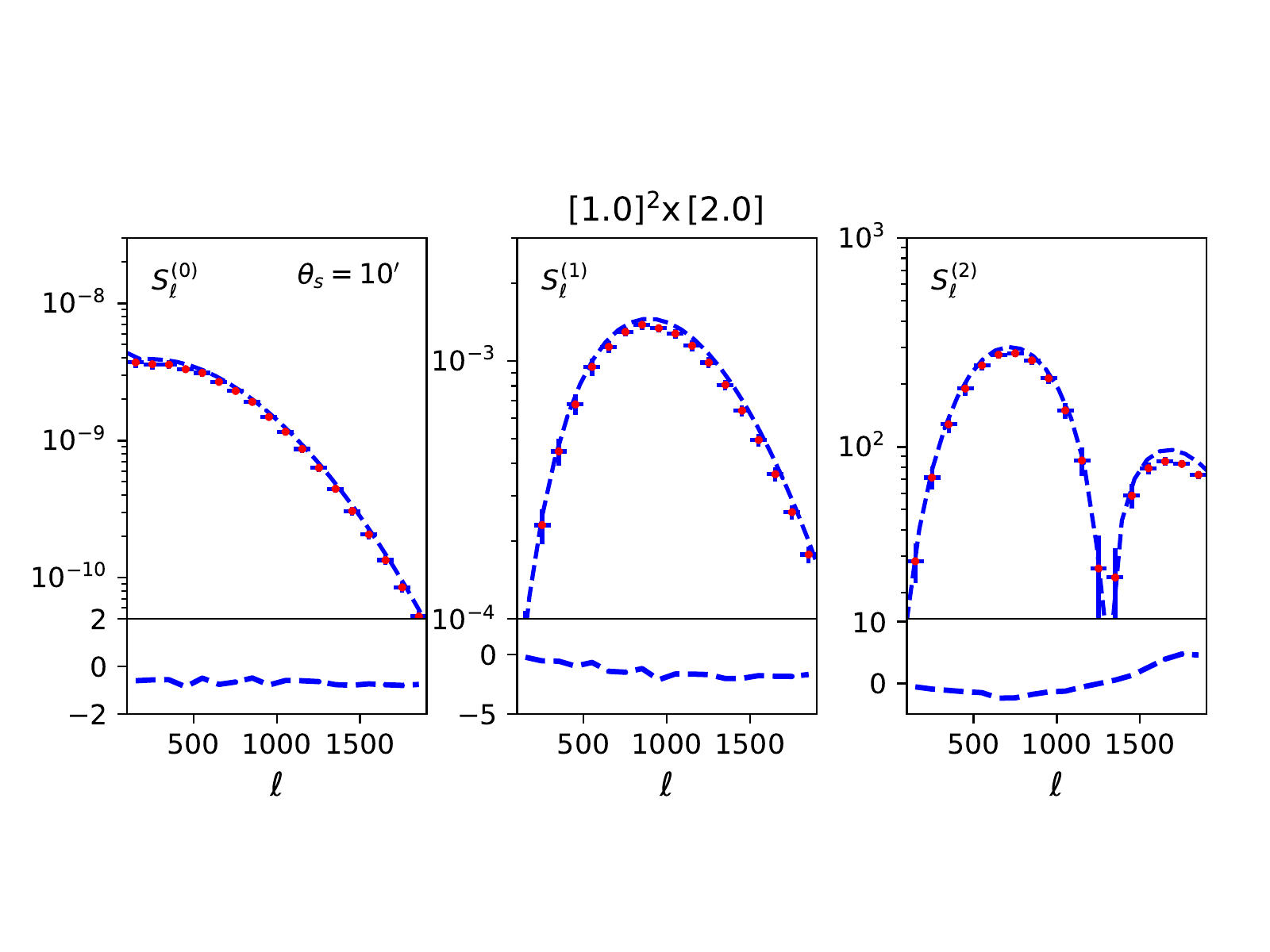}
  \vspace{-1.15cm}
  \hspace{1.cm}
  \caption{We have chosen two redshift bins $z_1=1.0$ and $z_2=2.0$. From left to right we show results for
    the skew-spectra $S^{(0)}_{\ell}$, $S^{(1)}_{\ell}$ and  $S^{(2)}_{\ell}$ as a function of
    $\ell$. The smooth curves represent theoretical predictions where as data points represent estimates
    from the simulations. These generalised skew-spectra are defined in Eq.(\ref{eq:s0})-Eq.(\ref{eq:s2}). In each panel
    we show $\langle \kappa_1^2\kappa_2 \rangle$ (in our notation, $\kappa_1 = \kappa(z_1)$) and $\kappa_2=\kappa(z_2)$)
    and $\langle \kappa_1^2\kappa_2 \rangle$ for two different smoothing
    angular scales $\theta_s=10'$. One single all-sky map was used to construct the skew-spectra.
    No noise was included in our study.}
  \label{fig:cross1}
\end{figure}

\begin{figure}
  \centering
  \includegraphics[width=0.7\textwidth]{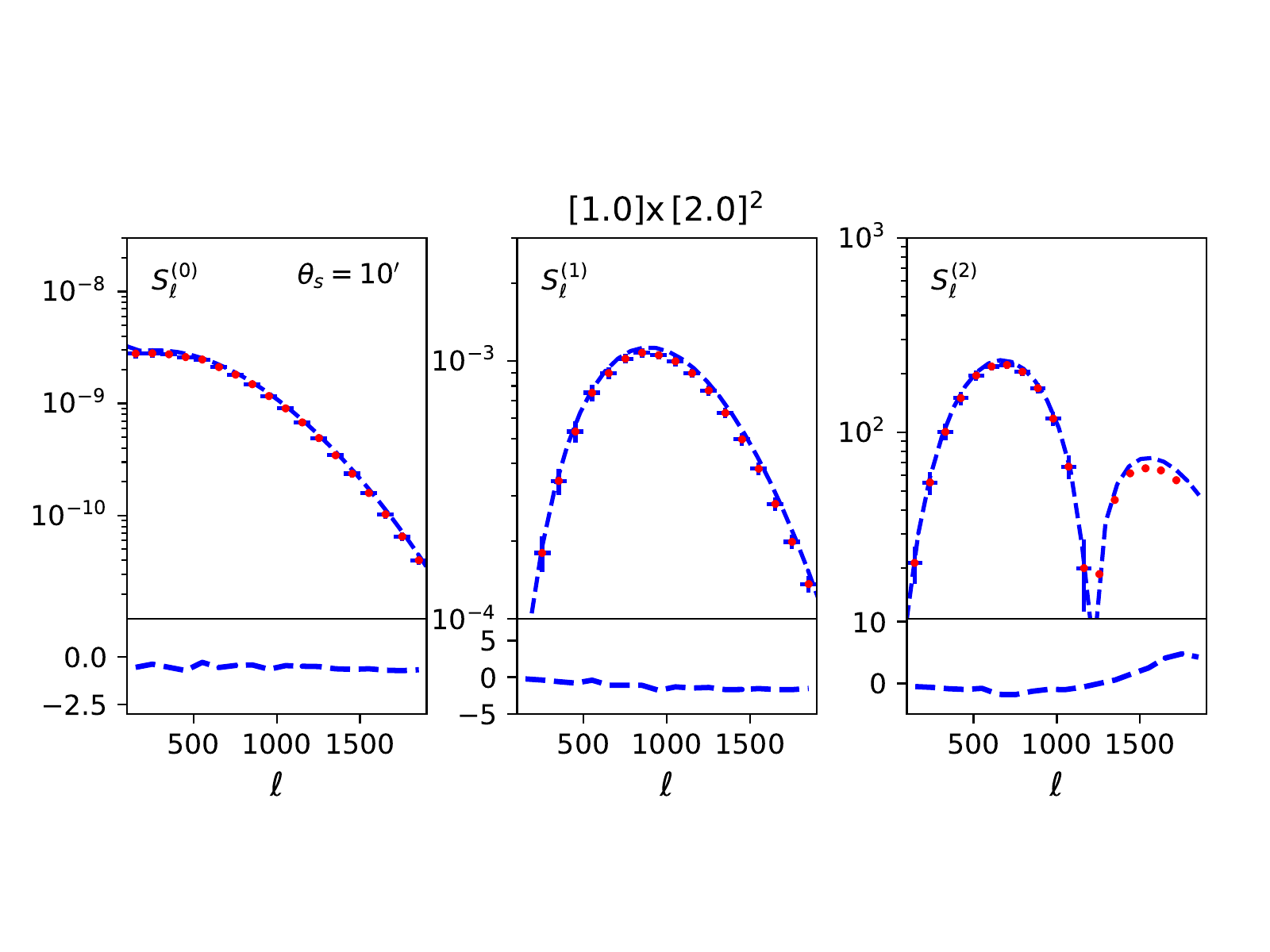}
  \vspace{-1.5cm}
  %\hspace{1.cm}
  \caption{Same as Figure-\ref{fig:cross1} but the skew-spectra associated with $\langle\kappa_1\kappa_2^2\rangle$
  is being plotted.}
  \label{fig:cross4}
\end{figure}

\begin{figure}
  \centering
    %\begin{minipage}[b]{0.99\textwidth}
  \includegraphics[width=0.7\textwidth]{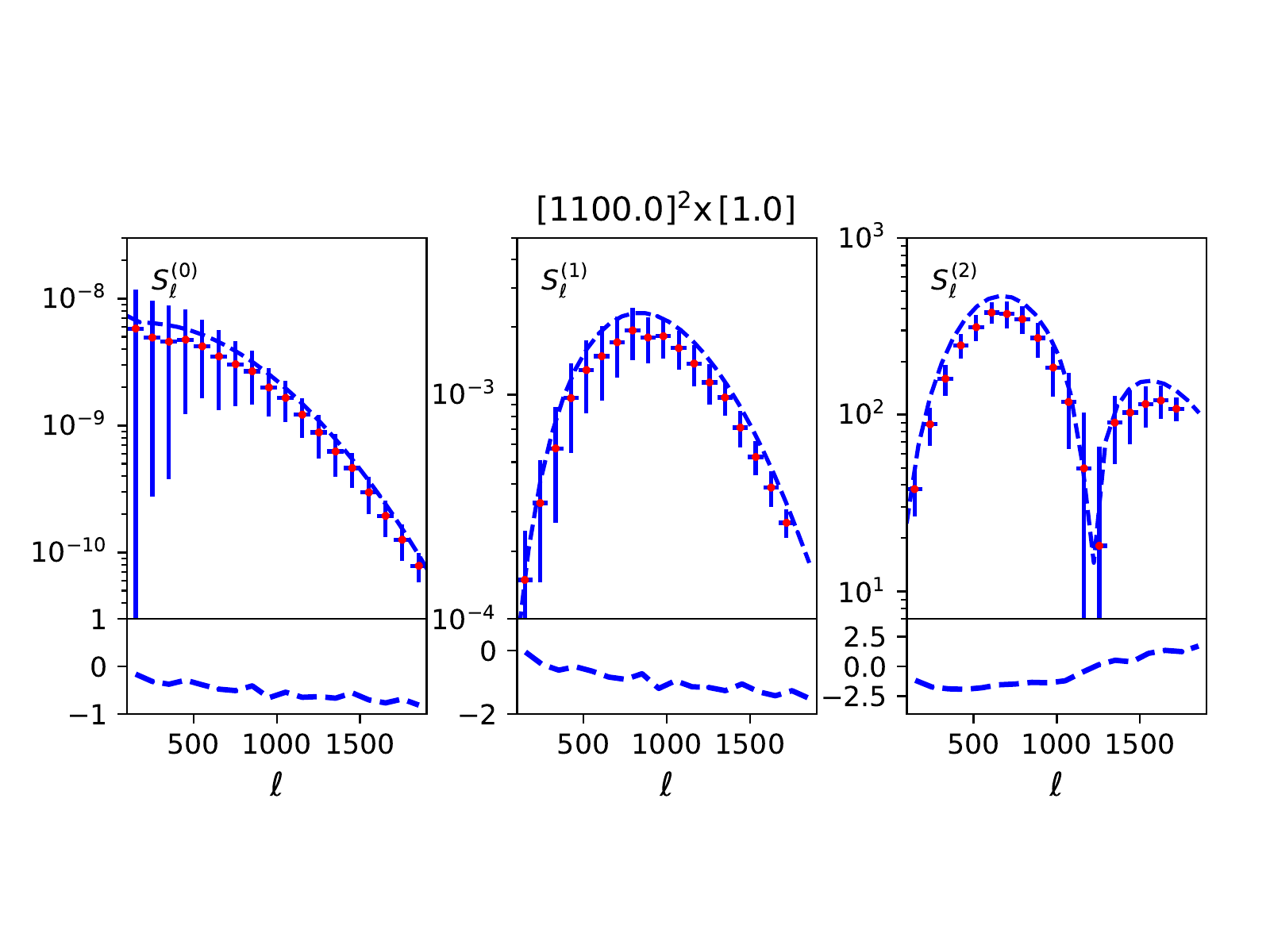}
  \vspace{-1.25cm}
  \caption{Same as Figure-\ref{fig:cross1} but for $z_1 =1100$ and $z_2 =1.0$.
  For $z_s=1100$ the $\kappa$ is being inferred from CMB observations.}
  \label{fig:cross2}
\end{figure}

\begin{figure}
  \centering
  \includegraphics[width=0.7\textwidth]{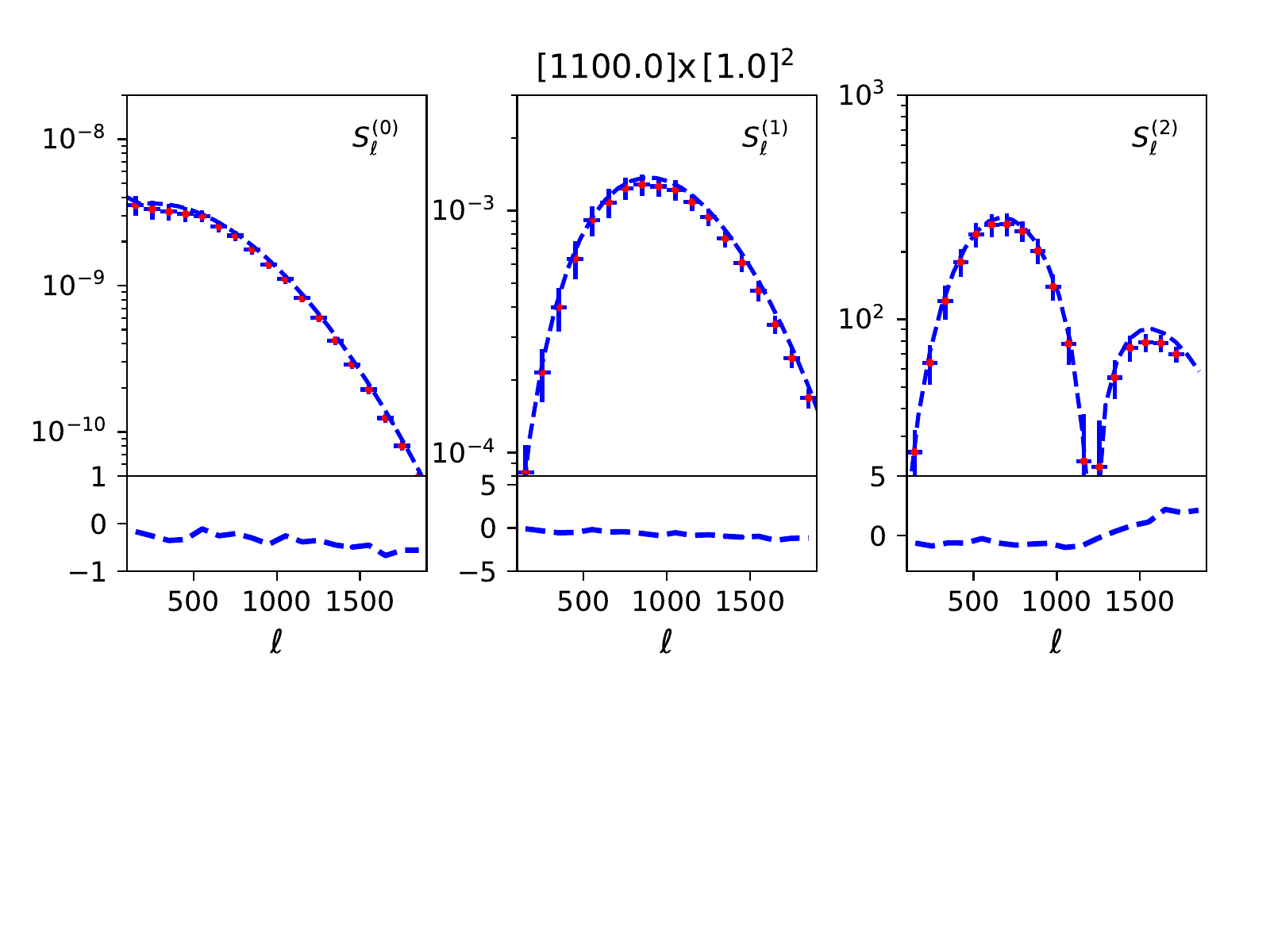}
  \vspace{-2.5cm}
  %\hspace{1.cm}
  \caption{Same as Figure-\ref{fig:cross2} but for $z_1=1.0$ and $z_2=1100.0$.}
  \label{fig:cross3}
\end{figure}

In this section we will summarize the main results presented in this paper along with their implications.
\begin{enumerate}
\item  
  {\bf Skew-spectra for individual tomographic bins at a low redshift:} 
  In Fig.-\ref{fig:save_mink1} --\ref{fig:save_mink4}
  the generalised skew-spectra $\essA$, $\essB$ and $\essC$  (from left to right) are being plotted as a
  function of $\ell$. These figures correspond to different source redshifts $z_s=0.5$, $z_s=1.0$, $z_s=1.5$ and $z_s=2.0$
  respectively. The various line styles in each panels correspond to different smoothing angular scales.
  We use a Gaussian window in our study. From top to bottom different curves represent
  Full Width at Half Maxima (FWHM) of $\theta_s=2.0'$, $\theta_s=5.0'$ and $\theta_s=10.0'$ respectively.
  We use the noise free simulations described in \ref{sec:simu}. We have used Eq.(\ref{eq:s0})-Eq.(\ref{eq:s2}) to evaluate the
  theoretical expectations for $\essA$, $\essB$, $\essC$ along with the fitting function by \citep{Ryuichi} discussed
  in \textsection\ref{sec:fit}. We have used theoretical predictions with and without the post-Born approximation
  but we find inclusion of such corrections make no significant impact on theoretical predictions.
  Over the entire range of smoothing angular scales $\theta_s$ and angular harmonics $\ell$ studied
  we haven't found any significant departure from theoretical predictions.
  We have used $N_{\rm side}=4096$ in our study. The skew-spectra are sensitive to the $\ell_{max}$.
  We have included all modes up to $\ell_{max} = 2000$ in our calculation in our theoretical predictions.
  To be consistent we have also filtered all modes higher than $\ell_{max}$ while processing
  the numerical simulations. We have also tested the impact of retaining the lower harmonics in our
  numerical evaluation by filtering out these modes from the maps as well as keeping them
  in while computing the skew-$S_{\ell}$s. We didn't find any statistically significant
  difference in our final results. The flexibility and simplicity with which the skew-spectra can be
  evaluated gives a very efficient to study the spectra in a mode-by-mode manner thus providing
  a greater handle on dealing with any possible systematics.
  Notice that the perturbative reconstruction of the MFs 
  requires the expansion parameter $\sigma_0$ introduced in  Eq.(\ref{eq:edge}) to be small for the 
  series to be convergent but, the three skew-spectra can also be used as independent estimators
  of non-Gaussianity and a method of effective data compression in their own right.
  This makes them attractive even when the series in Eq.(\ref{eq:v_k}) is divergent at smaller angular scales.
  The convergence of the series expansion and its implications were considered in \citep{Petri} to some extent.
  However, a detailed study is needed for a realistic assessment as a function of various
  survey parameters.
\item
  {\bf Skew-spectra from CMB maps:} In Fig.-\ref{fig:cmb_skew1}-Fig.\ref{fig:cmb_skew3}
        the generalised spectra $S^{(0)}$  (left panel), $S^{(1)}$ (middle panel) and $S^{(2)}$ (right panel)
        are plotted for redshift $z_s=1100.0$. The convergence maps are inferred from CMB observations.
        The variance or skew-spectra increases with redshift or the depth of the survey.
        To reduce the scatter in our estimates we have used binning with bin-size $\Delta_\ell=100$.
        While in Fig.-\ref{fig:cmb_skew1} the smoothing angular scale is sized at $\theta_s=2'$, in
        Fig.-\ref{fig:cmb_skew2} and Fig.-\ref{fig:cmb_skew3} this angular scale is fixed respectively
        at $\theta_s=5'$ and $10'$. The dot-dashed lines correspond to Born approximation.
        The dot-dashed lines in each panel include the post-Born corrections.
        The important difference of the CMB skew-spectra with the ones at lower redshifts
        is the significance of post-Born correction in modelling of non-Gaussianity.
        The post-Born correction is non-linear and it is known to generate a non-negligible bispectrum
        of the convergence \citep{Marozzi16, Prattenlewis}.
        Our study confirms that the post-Born contributions to the bispectrum can significantly change the shape predicted
        for the skew-spectrum from the large-scale structure non-linearities alone. This is more obvious in the
        right panels where the generalised skew-spectrum $S^{(2)}_{\ell}$ changes a signature from positive at lower $\ell$ to
        negative at higher $\ell$.
\item
  {\bf Skew-spectrum from cross-correlating two different tomographic bins:}
  In addition to studying the skew-spectra from individual tomogrpahic bins we have also
  cross-correlated different bins to construct the skew-spectra. Indeed the link to
  morphology no longer exists but this gives us a clue about how these estimators are correlated.
  It can also be argued, irrespective of morphological connection, that these estimators 
  provide an efficient tool for data compression. 
  
  In Fig.-\ref{fig:cross1} and  Fig.-\ref{fig:cross4} we show the cross skew-spectra of
  two tomographic bins $z_s=1.0$ and $z_s=2.0$.
  We have fixed $\theta_s=10'$ in each of these plots. The error-bars are computed using
  the fluctuations within a bin. The bin size is $\Delta_{\ell}=100$. In each case we find that the
  analytical and numerical predictions agree within $2\sigma$ in the cosmic variance limited case.

  In Fig.\ref{fig:cross2} and Fig.\ref{fig:cross3} we plot the skew-spectra constructed from
  $\kappa$ maps inferred from CMB observations at $z_s=1100$ (denoted as $\kappa_{\rm LSS}$) and cross-correlated
  against convergence map at $z_s=1.0$ (denoted as $\kappa_1$). In
  Fig.\ref{fig:cross2} we plot the skew-spectra related to $\langle\kappa_{\rm LSS}^2\kappa_1\rangle$ and
  in Fig.\ref{fig:cross3} the skew-spectra corresponding to $\langle\kappa_{\rm LSS}\kappa_1^2\rangle$ is being plotted.
  Compared to the low-$z$ cases the theoretical predictions for $\langle\kappa_1 \kappa^2_{\rm LSS}\rangle$ are found to significantly
  over-estimate the simulation results. This is true to a lesser extent for $\langle\kappa_1^2\kappa_{\rm LSS}\rangle$.
  This may be related to the fact that the simulation using a Gaussian realisations at higher redshifts $z_s >7.1$
  which may lead to suppression of non-Gaussinity. The descrepency becomes, however, not so significant when compared with
  the scatter within the beam.
\item
  {\bf Euclid-like Mask, Noise and Skew-spectrum:} In Figure-\ref{fig:Euclid} we show the three skew-spectra for a Euclid-like survey.
  We use a
    “pseudo {\em Euclid}” mask. To construct this mask all pixels lying within $22 \deg$ of either the galactic
    or ecliptic planes are discarded. Such a mask leaves  
    $14, 490 \deg^2$ of the sky making i.e. fraction of the sky covered
    $f_{\rm sky} ≈ 0.35$ (see \citep{Skew1} for more detailed discussion).
    We use maps with source plane fixed at $z_s = 1.0$.
    In each panel the upper curves correspond to the all-sky $S_{\ell}$ estimates and
    the lower curves correspond to the pseudo-$\hat S_{\ell}$s (see Eq.(\ref{eq:PCL1})). To compute the scatter one realization of the map was considered.
    To simulate noise we have included a source density of $n_s = 30\, {\rm arcmin^{-2}}$. However, we found that the Euclid-type noise
    do not produce any significant effect on the scatter. To increase the effect of noise we have artificially increased
    the level of noise by a factor of two.

\end{enumerate}
%
%%%%%%%%%% Figure -3 %%%%%%%%%%%%%%%%%%%%%%%
%%%%%%%%%%%%%%%%%%%%%%%%%%%%%%%%%%%%%%%%%%%%%
%
%%%%%%%%%% Figure -3 %%%%%%%%%%%%%%%%%%%%%%%%%
%%%%%%%%%%%%%%%%%%%%%%%%%%%%%%%%%%%%%%%%%%%%%
%
%%%%%%%%%% Figure -3 %%%%%%%%%%%%%%%%%%%%%%%%%
%%%%%%%%%%%%%%%%%%%%%%%%%%%%%%%%%%%%%%%%%%%%%
%
%%%%%%%%%% Figure -3 %%%%%%%%%%%%%%%%%%%%%%%%%
%%%%%%%%%%%%%%%%%%%%%%%%%%%%%%%%%%%%%%%%%%%%%
%
%
%%%%%%%%%%%%%%%%%%%%%%%%%%%%%%%%%%%%%%%%%%%
\section{Conclusions and Future Prospects}
\label{sec:conclu}
%%%%%%%%%%%%%%%%%%%%%%%%%%%%%%%%%%%%%%%%%%
%
%
%%%%%%%%%% Figure -3 %%%%%%%%%%%%%%%%%%%%%%%%%
\begin{figure}
  \centering
  \includegraphics[width=0.7\textwidth]{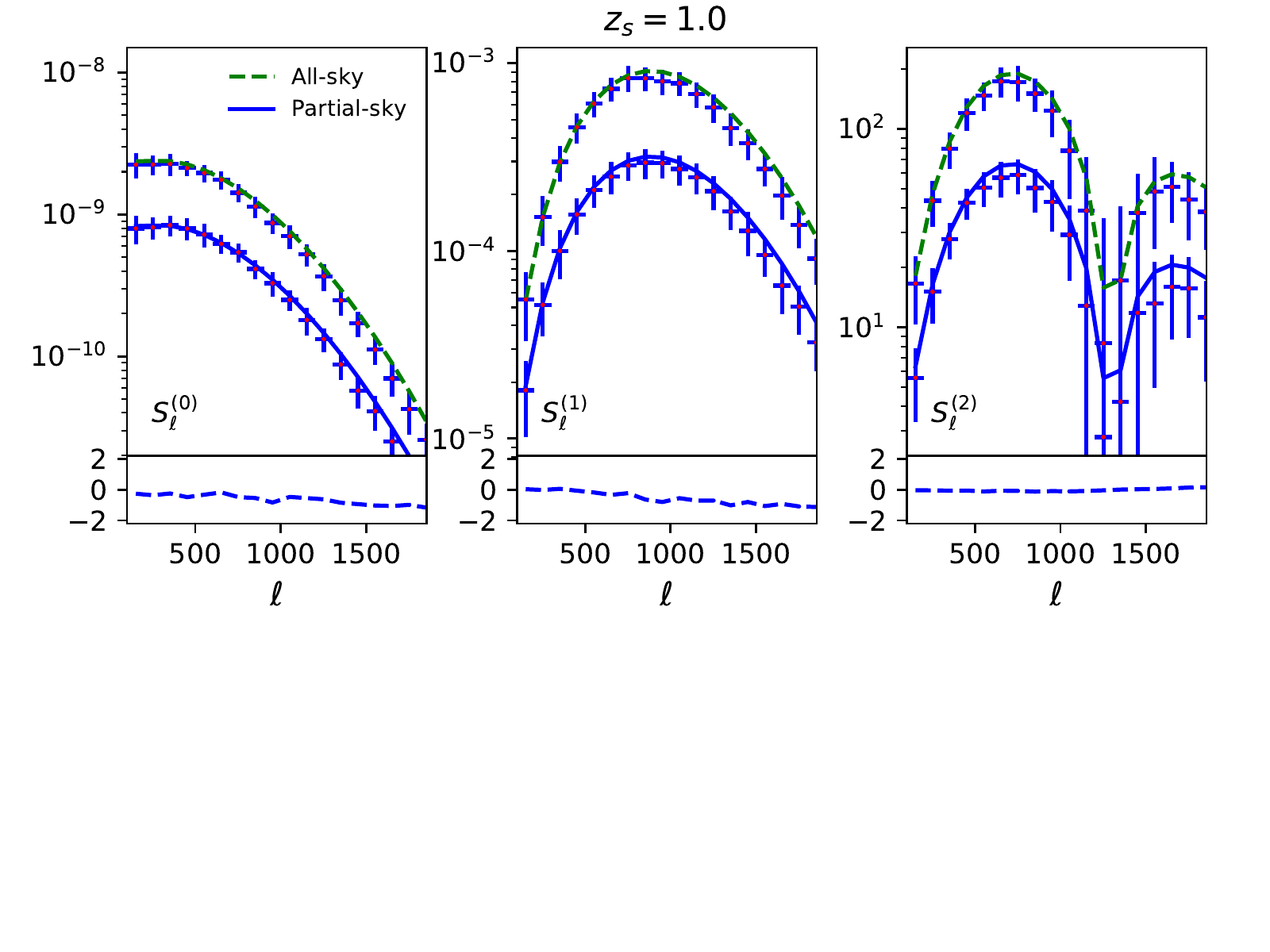}
  \vspace{-3.cm}
  \caption{We show the three skew-spectra for a Euclid-like survey. In our study we use a
    “pseudo Euclid” mask. All pixels lying within $22 \deg$ of either the galactic
    or ecliptic planes are discarded while constructing the mask. Which leaves  
    $14, 490 \deg^2$ of the sky making fraction of the sky covered $f_{\rm sky} ≈ 0.35$
    (see \citep{Skew1} for more detailed discussion).
    The source plane is fixed at $z_s = 1.0$. In each panel the upper curves correspond to the all-sky $S_{\ell}$ estimates and
    the lower curves correspond to the pseudo-$\hat S_{\ell}$s (see Eq.(\ref{eq:PCL1})). One realization of the all-sky maps were considered.
    To simulate noise we have included a source density of $n_s = 30\, {\rm arcmin^{-2}}$. With Euclid type noise the
    error-bars are nearly identical to what was presented in Figure-\ref{fig:save_mink2}. To amplify the
  effect of noise we have artificially increased the noise by a factor of two.}
  \label{fig:Euclid}
\end{figure}
%#############################################
%
%
The high signal-to-noise of the skew-spectra and
the flexibility with which they can be implemented
is rather encouraging. The accuracy of
the fitting function in reproducing
the numerical simulations opens up 
several possible avenues of research. 

{\bf Perturbative contributions from trispectrum:}
Beyond the leading-order non-Gaussian corrections, that come from bispectrum,
the four generalised kurtosis parameters  $K^{(0)}$, $K^{(1)}$, $K^{(2)}$, and $K^{(3)}$,
play an important role in perturbative reconstruction of
the morphology of a non-Gaussian field. These are the contributions
denoted as $\delta V^{(3)}_k$ in Eq.(\ref{eq:edge}).
These kurtosis parameters were generalised to kurtosis-spectra in a manner similar to the generalisation
of the skewness parameters to the skew-spectrum \citep{MuHu16}. The kurtosis-spectra were 
used in the context of CMB studies and sources
of non-Gaussianity studied include the primordial non-Gaussianity 
as well as lensing induced non-Gaussianity. 
Extension of our results to incorporate higher-order terms
in the context of weak lensing studies for gravity induced
non-Gaussianity will require an analytical model of the
trispectrum. The analytical expression for the  perturbative trispectrum
is more involved and will require a dedicated study.
Various other options to include the validity domain of the
perturbative expression include Effective Field Theoretic (EFT)
or Halo Model (HM) based approaches. We plan to extend our results
in future in these directions. 

{\bf Study of morphology from shear maps:} In our study we have extracted the generalised skew-spectra
directly from convergence maps. This requires an intermediate step of map making
from shear maps. However, our method can also be generalised to
directly deal with shear maps by implementing an Electric/Magnetic ($\rm E/B$)
decomposition of shear maps. The PSL approach can be generalised to
deal with such a decomposition and deal with arbitrary mask.
This will be useful in bypassing the map making process needed
for generating convergence maps. This will also be
important dealing directly with spurious magnetic or $B$ mode generated
due to unknown systematics.

{\bf Likelihood Analysis and Covariance Matrix:}
Any cosmological parameter inference using MFs would require a detailed
characterization of covariance matrix of the skew-spectra.
The calculation of covariance matrices were presented in \citep{Waerbeke}
using a simplistic approach that is valid in the noise dominated regime
i.e. in the limit of vanishing non-Gaussianity. This is achieved by ignoring the contributions
from all higher-order non-Gaussianity. While such approximate treatment
may be enough to deal with present generation of surveys, stage-IV observation
including the {\em Euclid} will map the sky with higher signal-to-noise
and may require a more accurate modelling is thus required.

{\bf Intrisic Allignment:} The intrinsic alignment (IA) remains a major contamination
to the gravity induced secondary non-Gaussianity. Analytical modelling of
IA is challenging though quite  a few physically motivated models can
capture certain aspects of the non-Gaussianity induced by IA \citep{VCS10}.
Typically at the level of bispectrum, IA is expected to contribute at $10\%$ of
the gravity induced non-Gaussianity. Using the skew-spectra introduced
here it will be possible to compute the corrections to the
morphological change induced by IA. In addition
optimal weights combined with a match filtering approach can in
effect may lead to separation of the two sources.

{\bf Betti number and other topological estimators:}
The MF were recently generalised in a series of paper
to Tensorial Minkowski Functionals (TMF) in 2D and 3D as well as in redshift-space
\citep{Tense1,Tense2}. The results presented here will be extended to the
case of TMF for a 3D convergence maps in future. Other estimators related to morphology
of cosmological fields have recently attracted attention, such as the Betti numbers \citep{Pranav19}. Reconstruction
techniques used here can be useful in these contexts.

{\bf Optimality and Flexibility of implementation:} We have not included optimal weighting in our estimator
  as the signal-to-noise is very high for low source redshift studies. This is not completely true
  for the studies involving $\kappa$ maps. Various methods can be used to improve
  the signal-to-noise including a Wiener or ``Wiener-like'' filtering of $\kappa$ maps \citep{Bouchet13}.
  Alternatively following \citep{MunshiHeavens} the generalised skew-spectra
  can include optimal weights that inherits a match filtering approach.
  However, there is a price to pay as the direct links to morphology will be lost and the estimators
  will have less flexibility in dealing with partial sky coverage as the PSL developed in our
  study will not be valid.

{\bf Beyond $\Lambda$CDM scenarios :}
  Though we have only discussed the gravity induced secondary non-Gaussianity as a possible
  source of non-Gaussianity, many other source of non-Gaussianities can also be
  included in our framework e.g. primordial non-Gaussianity or non-Gaussianity
  induced by active source of perturbations or topological defects can also be studied
  using their impact on morphology of convergence maps. Many modified gravity
  theories predict a different form of bispectrum compared to General Relativity and
  their impact on morphology can be studied using the formalism developed here \citep{MunshiMcEwen20}.

%%%%%%%%%%%%%%%%%%%%%%%%%%
\section*{Acknowledgment}
%%%%%%%%%%%%%%%%%%%%%%%%%%
%
DM is supported by a grant from the Leverhume Trust at MSSL.
DM would like to thank Chiaki Hikage and Geraint Pratten for useful discussions during the
initial phase of this project. DM would also like to thank the members of Euclid
Forward Modelling Working group including Benjamin Wandelt, Adam Amara and Martin Kilbinger for critical comments.
We would like to thank Peter Taylor for providing us his code to generate
the Euclid type mask used in our study. 
\bibliography{sub_mink.bbl}
\end{document}